\documentclass[10pt,journal,compsoc]{IEEEtran}

\ifCLASSOPTIONcompsoc
  \usepackage[nocompress]{cite}
\else
  \usepackage{cite}
\fi

\ifCLASSINFOpdf

\else

\fi

\usepackage{amsmath}
\usepackage{amssymb}
\usepackage{chemarrow}
\usepackage{balance}
\usepackage{hyperref}
\usepackage{amsmath} 
\usepackage{graphicx}
\usepackage{subfigure}
\usepackage{url}
\usepackage{color}
\usepackage{breqn}
\newcommand{\BfPara}[1]{{\noindent\bf#1.}\xspace}
\usepackage{multirow}
\usepackage{float}
\usepackage[export]{adjustbox}
\usepackage[T1]{fontenc}
\usepackage[inline]{enumitem}
\usepackage[utf8]{inputenc}
\usepackage{environ}
\usepackage{tikz}
\usetikzlibrary{arrows}
\usetikzlibrary{calc,matrix}
{\bfseries}{\itshape}
{\bfseries}{\itshape}
{\bfseries}{\itshape}
{\bfseries}{\itshape}
{\bfseries}{\itshape}
{\itshape}

\usepackage{makecell}

\usepackage{xspace}
\usepackage{setspace}
\newcommand{\note}[1]{}

\newcommand{\ie}{{\em i.e.,}\xspace}
\newcommand{\cc}{{cryptocurrency}\xspace}

\newcommand{\ep}{{{\em e-PoS}}\xspace}

\newcommand{\tsref}[1]{\textsection\ref{#1}\xspace}

\newcommand{\etal}{{\em et al.}\xspace}

\newcommand{\sign}{\mathsf{Sign}}
\newcommand{\keygen}{\mathsf{KeyGen}}
\newcommand{\verify}{\mathsf{Verify}}

\usepackage{cite}
\usepackage{hyperref}
\definecolor{linkcolour}{rgb}{0,0.2,0.6}
\definecolor{xgreen}{rgb}{0.2,0.6,0.0}
\definecolor{xred}{rgb}{0.7,0.1,0.0}
\hypersetup{colorlinks,breaklinks,citecolor=black, urlcolor=black, linkcolor=black}
\def\equationautorefname~#1\null{(#1)\null}

\usepackage[capitalise]{cleveref}

\usepackage[linesnumbered,ruled,noend]{algorithm2e}

\colorlet{punct}{red!60!black}
\definecolor{background}{HTML}{ffffff }
\definecolor{delim}{RGB}{20,105,176}
\colorlet{numb}{magenta!60!black}

\newcommand*\circled[1]{\tikz[baseline=(char.base)]{%
            \node[shape=circle,fill=black,draw,text=white,inner sep=0.5pt] (char) {#1};}}

\definecolor{light-gray}{gray}{0.95}
\definecolor{darkgray}{rgb}{0.4, 0.4, 0.4}
\definecolor{editorGray}{rgb}{0.95, 0.95, 0.95}
\definecolor{editorOcher}{rgb}{1, 0.5, 0} 
\definecolor{editorGreen}{rgb}{0, 0.5, 0} 
\definecolor{orange}{rgb}{1,0.45,0.13}      
\definecolor{olive}{rgb}{0.17,0.59,0.20}
\definecolor{brown}{rgb}{0.69,0.31,0.31}
\definecolor{purple}{rgb}{0.38,0.18,0.81}
\definecolor{lightblue}{rgb}{0.1,0.57,0.7}
\definecolor{lightred}{rgb}{1,0.4,0.5}
\usepackage{upquote}
\usepackage{listings}

\definecolor{pblue}{rgb}{0.13,0.13,1}
\definecolor{pgreen}{rgb}{0,0.5,0}
\definecolor{pred}{rgb}{0.9,0,0}
\definecolor{pgrey}{rgb}{0.46,0.45,0.48}

\makeatletter
\let\matamp=&
\catcode`\&=13
\makeatletter
\def&{\iftikz@is@matrix
  \pgfmatrixnextcell
  \else
  \matamp
  \fi}
\makeatother

\newcounter{lines}

\newcounter{vtml}
\hypersetup{
	pdfpagemode=pagewidth,
	plainpages=false,
	colorlinks,
	urlcolor=blue!70!black,
	linkcolor=red!70!black,
	citecolor=green!70!black,
	bookmarksnumbered
}

\begin{document}

\title{\ep: Making Proof-of-Stake \\ 
 Decentralized and Fair}

\author{Muhammad Saad,
        Zhan Qin, Kui Ren, DaeHun Nyang,
        and David Mohaisen
\IEEEcompsocitemizethanks{\IEEEcompsocthanksitem Muhammad Saad and David Mohaisen are affiliated with the University of Central Florida, USA. Zhan Qin and Kui Ren are affiliated with the Zhejiang University, China. DaeHun Nyang is with the the Ewha Womans University. South Korea. 

E-mail: saad.ucf@knights.ucf.edu, qinzhan8871@zju.edu.cn, kuiren@zju.edu.cn, nyang@ewha.ac.kr, and mohaisen@ucf.edu. David Mohaisen is the corresponding author. This work was supported in part by NRF grant NRF-2016K1A1A2912757 (GRL Project).}

}

\markboth{IEEE Transactions on Parallel and Distributed Systems \ }
{\MakeLowercase{\textit{}}}

\IEEEtitleabstractindextext{%
\begin{abstract}
Blockchain applications that rely on the Proof-of-Work (PoW) have increasingly become energy inefficient with a staggering carbon footprint. In contrast, energy efficient alternative consensus protocols such as Proof-of-Stake (PoS) may cause centralization and unfairness in the blockchain system. To address these challenges, we propose a modular version of PoS-based blockchain systems called \ep that resists the centralization of network resources by extending mining opportunities to a wider set of stakeholders. Moreover, \ep leverages the in-built system operations to promote fair mining practices by penalizing malicious entities. We validate \ep{}'s achievable objectives through theoretical analysis and simulations. Our results show that \ep ensures fairness and decentralization, and can be applied to existing blockchain applications. 
\end{abstract}

}

\maketitle

\IEEEdisplaynontitleabstractindextext

\IEEEpeerreviewmaketitle

\ifCLASSOPTIONcompsoc

\section{Introduction}\label{sec:introduction}
Blockchain systems use various leader election schemes for consensus-driven ledger updates. In the Proof-of-Work (PoW) applications, for example, the protocol requires the network entities to produce a valid solution to a challenge to be able to publish an acceptable block. To avoid concurrent solutions, the challenge complexity is periodically calibrated to comply with the dynamically changing hash rate. Instances of this protocol can be observed in Bitcoin, where the challenge includes producing a $\textsl{nonce}$ string which, when hashed with the block header, produces a lower value than the {\em Target} threshold set by the system. As a reward for the efforts, the block releases coins to the miner's account. Once a block is published, all its transactions are approved and the block is added to the blockchain.

In cryptocurrencies, miners use sophisticated hardware to increase their chances of solving proof-of-work. Mining hardware consumes significant electricity to find the correct hash value, and only one solution is considered valid per block. Therefore, the energy used by all the ``losing'' miners is wasted. To make matters worse, multi-home mining pools have been set up worldwide to facilitate collaborative mining. These mining pools are similar to data center networks, with their own mining rigs, power houses, and cooling systems. Such processing-intensive operations expend high units of electricity. Bitcoin alone consumes more than 70 Terawatt-hour per year, which is more than the electricity consumed by countries, such as Colombia and Switzerland \cite{DEVRIES2018801,Digiconomist18}. The Bitcoin community has been criticized for enabling the exhaustion of valuable energy resources, and has been encouraged to adapt to green mining practices \cite{JacquetM18,XuWLGLYG19}.

An energy-efficient consensus alternative is known as the Proof-of-Stake (PoS), which relies on the stake of participants than their processing power. In PoS, miners are selected based on their balance (stake), and the network conforms to the view of those miners. In most PoS-based cryptocurrencies, coins are generated at the inception of the cryptocurrency itself. Therefore, there are no block rewards for miners. Instead, miners earn by collecting fee from transactions.  The selection of miners in PoS is similar to an auction process whereby miners bid for the next block and the miner who places the highest bid wins the auction~\cite{DeuberDMMT18,JiaoWNS19}. However, there is a major caveat in this scheme: a rich miner is likely to win more auctions, which in return makes this miner richer and increases his chances of winning subsequent auctions. Therefore, in the long term, PoS-based applications tend to become more centralized in favor of a few miners. Another challenge in PoS-based protocols is the lack of $\textsl{fairness}$ policies when system is under attack. For instance, if an adversary carries out fraudulent activities, such as double-spending, blockchain fork, or a majority attack, the existing systems do not have policies to penalize the adversary and compensate the victims.

To address the aforementioned challenges, we introduce a variant of PoS, called \ep, that resists network centralization, and promotes fair mining.We revisit the notions of decentralization and fairness to set baseline conditions to be  met. While decentralization guarantees that the auction process in \ep is made available to wider set of stakeholders, fairness ensures that, during an attack, malicious entities are penalized and the affected parties are compensated.

\BfPara{Contributions and Roadmap} Towards the broader goal of decentralization and fairness in PoS-based blockchain systems, we make the following key contributions: 
\begin{enumerate}[label=\protect\circled{\arabic*}, font=\small\bfseries ]
    \item We propose a consensus protocol, called the extended PoS (\ep), which addresses the limitations of PoW and PoS and enables decentralized, energy-efficient, and fair mining in blockchain systems.
    \item We present the design constructs of \ep and its translation into the blockchain framework using theoretical primitives and engineering requirements. 
    \item We perform feasibility and security analyses of \ep, and show its merits through simulations.
    \item Finally, to highlight structural and functional composability, we show how \ep can be applied to the existing cryptocurrencies such as Bitcoin and Ethereum. 
\end{enumerate}

Additionally, the paper includes background and motivation in section~\ref{sec:bg}, design primitives in section \ref{sec:pof}, simulations in section~\ref{sec:sm}, and concluding remarks in section~\ref{sec:con}.

\begin{figure}[t]
\hfill
\begin{subfigure}[Bitcoin\label{fig:crypto}]{\includegraphics[width=0.23\textwidth]{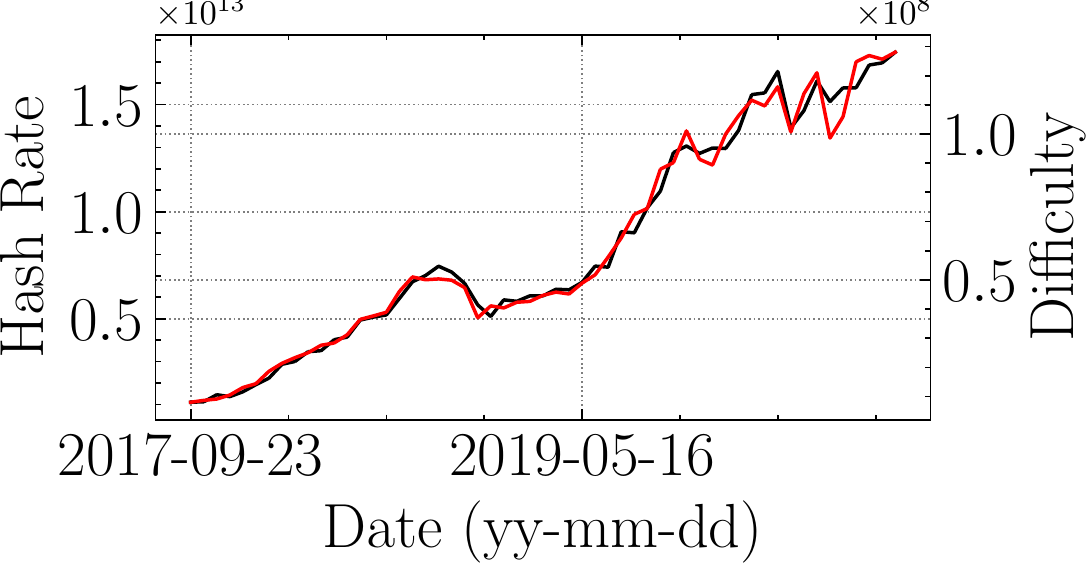}} 
\hfill
\end{subfigure}
\begin{subfigure}[Ethereum\label{fig:cryptotwo}]{\includegraphics[width=0.23\textwidth]{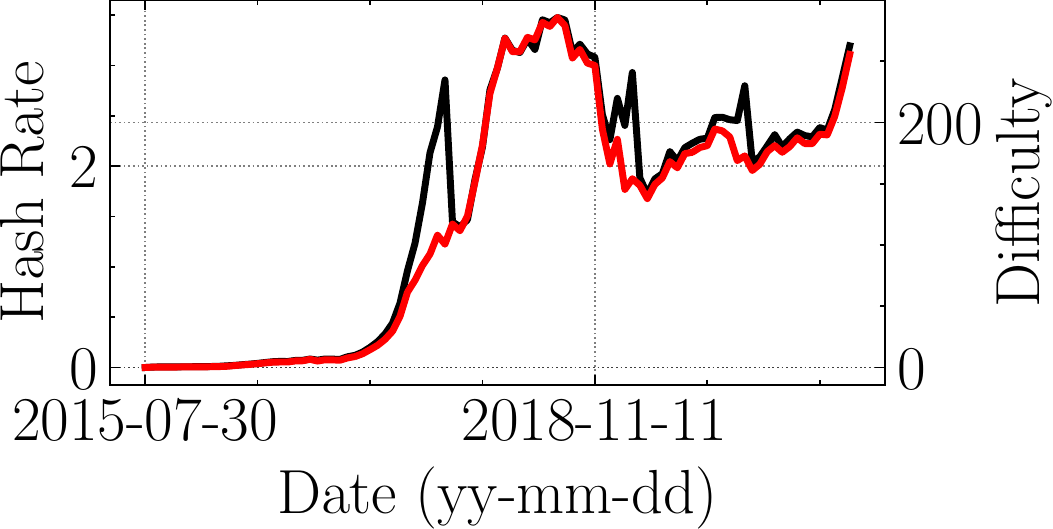}}
\end{subfigure}
\caption{\cref{fig:crypto} shows Bitcoin hash rate (EH/s) and difficulty, while ~\cref{fig:cryptotwo} shows Ethereum hash rate (PH/s) and difficulty.  }
\label{fig:nlabel}\vspace{-3mm}
\end{figure}

\section{Background and Motivation} \label{sec:bg}
\BfPara{Proof-of-Work Limitations} \label{sec:pow} 
PoW involves solving a mathematical challenge set by the network in order to produce a valid block. Now days, mining is carried out as a joint effort by the mining pools. As mining pools expand, the aggregate hashing power of the system increases, as shown in \cref{fig:crypto} and \cref{fig:cryptotwo}, and the competition for mining block increases as well. The excessive use of electricity in cryptocurrencies has raised major concerns. In 2019, Bitcoin consumed 69.97 TWh electricity with 32.95 Megaton carbon footprint~\cite{Digiconomist18}. The carbon footprint is the amount of carbon dioxide released due to electricity generation. Similarly, in 2019, Ethereum consumed 10.1 TWh electricity with 4.78 Megaton carbon footprint~\cite{digiconomist_20}. In~\cref{fig:ec} and \cref{fig:ectwo}, we present the energy profile of Bitcoin and Ethereum, respectively. Note that their energy profile is comparable to several countries, reflecting the threat they pose to the environment.

\BfPara{Proof-of-Stake Limitations} \label{sec:pos}
A popular alternative to PoW is the Proof-of-Stake (PoS), which overcomes the shortcomings of PoW while achieving similar objectives \cite{BadertscherGKRZ18}. In PoS, a candidate miner is selected to mine the next expected block, and the candidate miner has to put his balance as a stake. Peercoin, launched in 2012, is the first \cc that used PoS for block mining \cite{king2012ppcoin}. Later, more cryptocurrencies were launched including Nxt and BlackCoin, which used variants of PoS. The two popular techniques for miner selection are called the ``randomized block selection'' and the ``coin age-based selection.'' In the randomized block selection, a combination of the lowest hash value and the size of a candidate's stake is used to select the miner for the next block. In the coin age-based selection, an auction process selects the candidate miners based on the coin age.

\begin{figure}[t]
\hfill
\begin{subfigure}[Bitcoin\label{fig:ec}]{\includegraphics[width=0.23\textwidth]{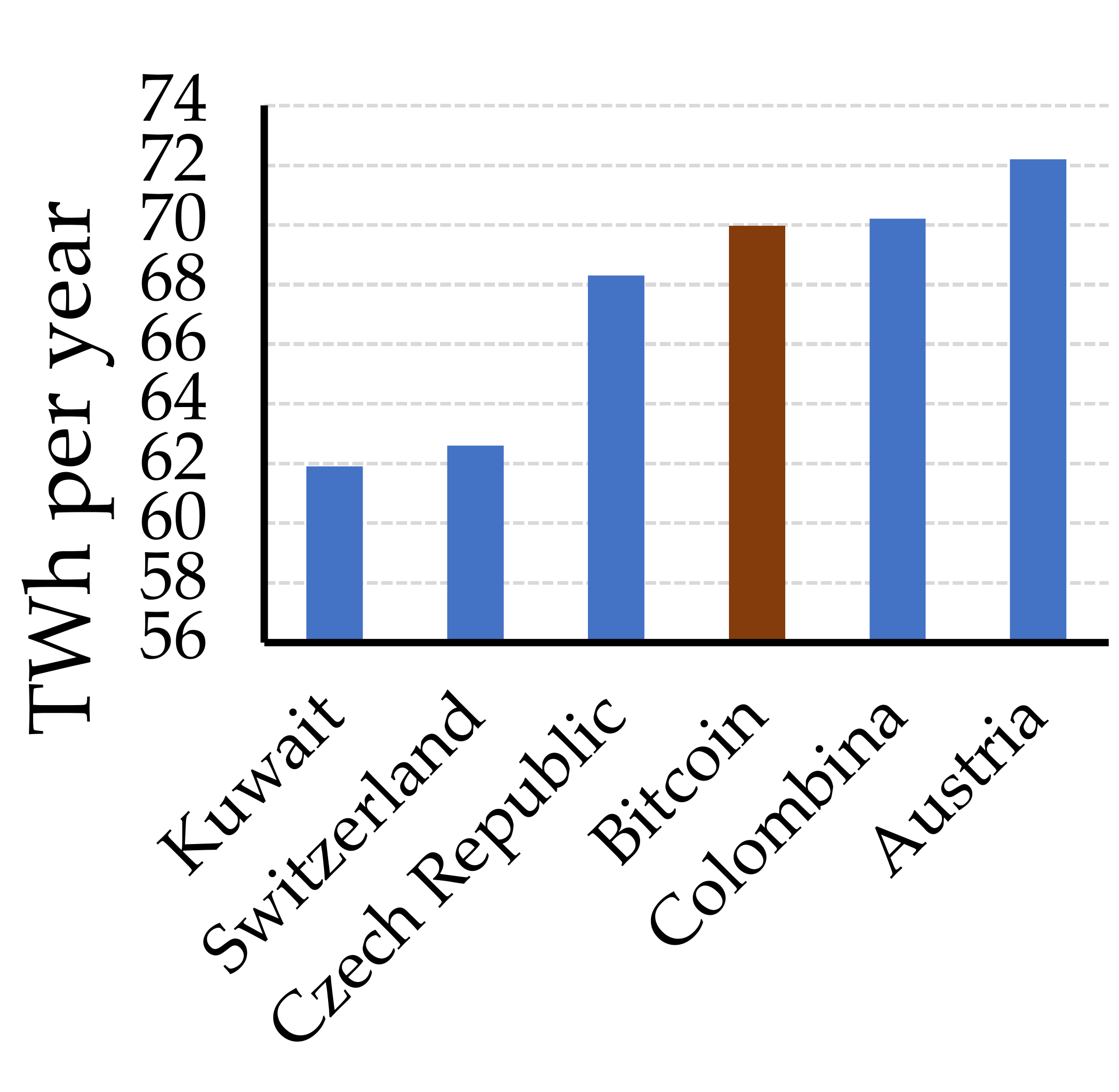}} 
\hfill
\end{subfigure}
\begin{subfigure}[Ethereum\label{fig:ectwo}]{\includegraphics[width=0.23\textwidth]{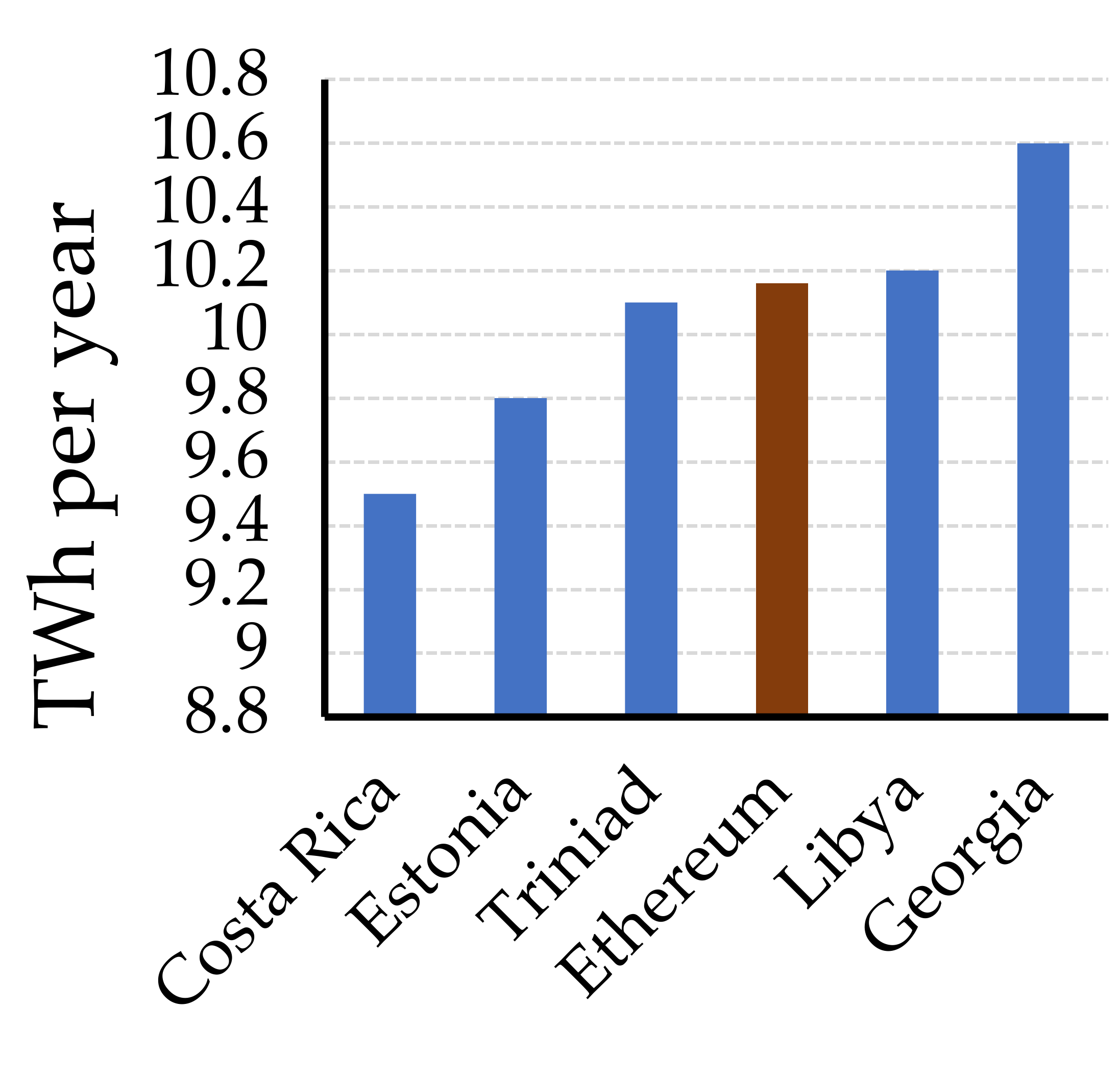}}
\end{subfigure}
\caption{Energy profile of Bitcoin and Ethereum. Note that Bitcoin's and Ethereum's electricity consumption is compare able to several  countries}
\label{fig:energy}\vspace{-3mm}
\end{figure}

Although PoS is energy-efficient, it introduces centralization~\cite{FantiKORVW18,Poelstra14,LiWYLZHLXD19}. In randomized block selection, the deserving candidates with high stakes are often ignored, and a random candidate is selected. In the open stake auction and age-based selection, a rich candidate is able to win every auction and get richer~\cite{buterin_2018, FantiKORVW18}. This creates a skew in the network, challenging the expected guarantees of the network decentralization, as generally expected in a blockchain application. Furthermore, public exposure of stakes reveals the balance of candidate miners, thereby compromising their anonymity and privacy. Moreover, in PoS, it is easy to fork the blockchain since anyone can produce a block without much $\textsl{effort}$ and fork the main chain. Such an attacker can exploit network latency and churn to identify nodes that lag behind the main chain, and partition them with his version of the blockchain~\cite{SaadCLTM19,SaadSNKSNM20}.

\BfPara{Motivation} \label{sec:moti}
Considering the aforementioned limitations, we anticipate a revision to the existing consensus schemes to improve decentralization and fairness. Naturally, we expect a variant of PoS or an energy-efficient hybrid model that does not rely on compute-intensive mining. For decentralization, the protocol should offer mining opportunities to a wider set of stakeholders. For fairness, the protocol must prevent fraudulent activities and fairly compensate affected parties if attacked. While ensuring fairness, the protocol must also have a reward mechanism that incentivizes honest mining. Additionally, it would be useful if the protocol complements the existing PoW-based cryptocurrencies to facilitate their transition to the proposed scheme and avoid bootstrapping complexities. With this motivation, we sketch the design of an algorithm that addresses these challenges and provides a blueprint for future applications.

\section{Extended Proof-of-Stake } \label{sec:pof}
Our protocol introduced in this section is called the ``Extended Proof-of-Stake'' (or \ep), which builds on PoS protocol, and offers fairness, security, and decentralization. .

\subsection{Design Overview} \label{sec:do}
The abstract implementation of \ep includes a set of miners executing an immutable smart contract that implements the rules of a PoS auction. The smart contract in \ep makes modular adjustments to the conventional PoS design to acquire the desirable properties including decentralization and fairness. Its correct execution, on the other hand, requires consensus among multiple parties (miners in a \cc). For that, we use the practical Byzantine fault tolerance (PBFT) protocol \cite{Castro00}. It can be argued that PBFT, in general, could provide an energy-efficient consensus scheme for distributed systems, although it suffers from high a message complexity which restricts its scalability for a \cc network comprising of thousands of nodes \cite{ChondrosKR11}. As we later show in \tsref{sec:app}, \ep fragments the \cc network in a way that the selected number of miners can easily execute PBFT. However, for now, we focus on the rules of the smart contract in \ep.

In \ep, the block mining is fragmented in a series of epochs $\mathcal{E}_{1},\ldots, \mathcal{E}_{j}$. In each epoch, the smart contract produces a sequence of blocks $\mathcal{B}_{1}, \ldots, \mathcal{B}_{l}$. The smart contract computes a baseline stake $\mathcal{ST}_{1}, \ldots, \mathcal{ST}_{l}$ for each block and announces them to the network. Candidate miners $\mathcal{C}(m_{c},b_{c})$, willing to participate in the block auction compare the baseline stake with their balance, and place a bid on their block of choice. The smart contract selects the final list of miners $\mathcal{K}(m_{k}, b_{k})$ based on their bids. For each epoch $\mathcal{E}_{j}$, the miners of the previous epoch $\mathcal{E}_{j-1}$ execute the smart contract using PBFT and transfer  control to the miners of the next epoch. Assuming a $3f+1$ honest majority among $\mathcal{K}(m_{k}, b_{k})$, \ep continues to function correctly~\cite{Castro00}.

\BfPara{Decentralization in \ep} \label{sec:decnetralization}
In PoS, decentralization can be attributed to the extension of mining opportunities to a wider set of stake holders~\cite{KwonLKSK19}. Roughly speaking, a PoS-based blockchain application is centralized if only the rich miners have the ability to mine a majority of the blocks.

For a given epoch of length $l$, let $n$ be the total number of participants in the network, $c<n$ be the number of candidate miners, and $k<c$, be the final list of miners in \ep. For simplicity, and without the loss of generality, we assume that two independent and concurrent execution fragments of PoS have been implemented, from which we obtain $k$  miners in \ep and $k'$ in the conventional PoS. Provided that, decentralization (or the lack of it, thereof) is expressed as a parameter $\beta$, where $0<\beta\leq 1$.  More precisely, $\beta_{e} = \frac{k}{n}$ is used for expressing the decentralization level in \ep, while $\beta_{c} = \frac{k'}{n}$ is used for expressing the decentralization in the conventional PoS. We note that a scheme is more decentralized if $\beta$ is closer to 1, and centralized otherwise. 

We use this notion of decentralization to compare different schemes. Let the difference between the two schemes above be $\gamma = \beta_{e} - \beta_{c}$, which represents the gap in decentralization between the two schemes. Such a gap is used to compare \ep the conventional PoS as follows:
\begin{align} \label{eq:decen}
    \beta_{e} - \beta_{c} &= \begin{cases} 
{\gamma<0} & \text{\ep is less decentralized}\\
{\gamma=0} &\text{equally decentralized} \\
{\gamma>0} &\text{\ep is more decentralized}
\end{cases}
\end{align}

The closer  $\gamma$ is to 1, the more decentralization \ep brings over the conventional PoS (and vice versa). To affirm that over a long run of the system, we say that one scheme is better than another (with respect to the decentralization notion) if the overwhelming majority of the observations of $\gamma$ over a large number of epochs is positive.

\BfPara{Fairness in \ep} \label{sec:fairness}
In blockchain applications, fairness is defined as a model that forces an adversary to pay back a penalty upon deviating from the protocol by, for example, premature abortion of the agreed upon protocol or by carrying out fraudulent activities. This notion of fairness is defined in the prior work to explore fair mining \cite{KiayiasZZ16,PassS17}. 

In general, an adversary may carry out fraudulent activities if that adversary has a significant advantage over other peers. In PoW, the advantage is the mining power. If the adversary holds more than 50\% hash rate, he can double-spend by mining a longer private chain and forcing the network to switch to the longer chain~\cite{nakamoto2008bitcoin}. In PoS, the same applies if the adversary holds more than 50\%  of all coins. In the following, we characterize this advantage of the adversary. Our objective is to show that the success probability \ep (with 50\% coins) is lower than in conventional PoS, thereby leading to a fair system with \ep. 

To formally analyze this aspect, let $B_N$ denote the total coins in the system, where $B_{N} = \sum_{i=1}^{n} b_{i}$. Let $\alpha B_{N}$ be the fraction coins owned by an adversary $\mathcal{A}$, and $\beta B_{N}$ be the coins owned by the honest miners. For $m$ consecutive blocks, the probability that the adversary successfully produces a longer private chain becomes:
\begin{align} \label{eq:adv2}
\text{\normalfont Pr} (~\mathcal{A}  \text{\normalfont ~double-spends in PoS})  &= \begin{cases} 
\bigg(\frac{\alpha}{\beta} \bigg)^{m}&, {\alpha<0.5} \\
1 &, {\alpha \geq 0.5}
\end{cases}
\end{align}
\begin{align} \label{eq:nadv2}
\text{\normalfont Pr} (~\mathcal{A}  \text{\normalfont ~mines the next block})  &= \begin{cases} 
\bigg(\frac{\alpha}{\beta}\bigg)&, {\alpha<0.5} \\
1 &, {\alpha \geq 0.5}
\end{cases}
\end{align}

\autoref{eq:adv2} shows that when $\alpha<0.5$, the adversary's success probability decreases exponentially. However, when $\alpha\geq 0.5$ ($\mathcal{A}$ acquires more than 50\% coins), the adversary eventually produces a longer private chain and double-spends. Moreover,~\autoref{eq:nadv2} shows that when $\alpha\geq 0.5$, the adversary mines the next block with probability 1. This is due to the fact that no other honest miner has a higher stake in the network. For any block, the adversary can place a higher bid than any other miner and win the auction. In summary, by acquiring more than 50\% coins, the adversary can (1) double-spend with probability 1, and (2) win each successive auction and mine subsequent blocks. This prevents honest miners from producing blocks and allows adversary to violate the blockchain safety properties~\cite{GarayKL17}. In other words, the adversary breaks the notion of fairness that we aim for a PoS-based blockchain system. In \ep, we propose that even with 50\% coins, the adversary's success probability of (1) double-spending is negligible (a small value $\epsilon$), and (2) mining the next block is less than 1. 
\begin{align}
\label{eq:adv3}\text{\normalfont Pr} (~\mathcal{A}  \text{\normalfont ~double-spends in \ep} ~|~\alpha\geq 0.5) =\epsilon\\
\label{eq:nadv3}\text{\normalfont Pr} (~\mathcal{A}  \text{\normalfont ~mines the next block in \ep} ~|~\alpha\geq 0.5) < 1
\end{align}
 
\subsection{Epoch Length} \label{sec:eps}
The first challenge in \ep is to determine the length of an eoch, $\mathcal{E}_{j}$, required to calculate the expected number of blocks $l$ to be computed. For that, we use the memory pool (mempool) of cryptocurrencies. The mempool is a repository of unconfirmed transactions used by miners to select transactions for blocks. The mempool size is usually greater than the average size of the block \cite{SaadTM18,SaadNKKNM19}. Miners prioritize transactions based on their fee, and select the high valued transactions from the mempool to mine in blocks.

The mining process can be viewed from two perspectives. 1) The miner's aim is to select highly paying transactions. 2) In ideal conditions, the network requires an empty mempool to prevent transaction backlog.

With a fixed block size, $\mathcal{B}_{s}$, and the size of mempool being $\mathcal{M}_{s}$, we can compute the number of blocks required to empty the mempool (i.e., to make $\mathcal{M}_{s} \approx 0$). It is desirable to completely fill the blocks with transactions such that the total size of each block must be equal to the standard block size of the cryptocurrency ($ |\mathcal{B}_{i}| \approx \mathcal{B}_{s}$). A complete block also means that the miner is able to obtain a maximum rewards from the mining fee paid by each transaction. In light of these design restrictions, the number of blocks in each epoch can be computed using \autoref{eq2}.
\begin{align} \label{eq2}
l= \begin{cases} 
{0} & \text{, $\mathcal{B}_{s} > \mathcal{M}_{s}$}\\
{\frac{\mathcal{M}_{s} }{\mathcal{B}_{s}}} &\text{, $\mathcal{B}_{s} < \mathcal{M}_{s}$}
\end{cases}. 
\end{align}
The number of blocks is computed by dividing the mempool size by the standard block size in 
\autoref{eq2}.

\BfPara{Epoch Duration} \label{sec:ed} The epoch duration $\mathcal{D}_{\mathcal{E}}$ is the product of the number of blocks produced in the epoch and the standard block time. Furthermore, the difference in the publishing time between two subsequent blocks should be equal to the standard block time to avoid undesirable delays in the block generation or the ledger update \cite{GarayKL17}. 

\subsection{Baseline Stake} \label{sec:bs}
In the next phase, the smart contract computes a baseline stake $\mathcal{ST}_{i}$ for each block $\mathcal{B}_{i}$ in the epoch. In light of the decentralization and fairness objectives outlined in \textsection\ref{sec:fairness}, the baseline stakes must satisfy the following principle objectives: 1) The baseline stakes must be small enough to allow mining opportunities to a large subset of miners ($c \approx n$). 2) The baseline stakes must also be large enough to prevent fraudulent activities and compensate victim parties. 3) The outcome of each mined block must proportionally benefit the candidate miners according to their commitments. 

Theoretically, the subset of candidate miners can be maximized by setting the baseline stakes to a negligible value ($\mathcal{ST}_{i} \simeq 0, \forall i$), such that the entire network becomes eligible for mining. However, this may challenge the two subsequent principle objectives that prevent fraud and promote incentivized reward distribution. The challenge is to construct a scheme that meets all three principle objectives. 

In \autoref{algo:sorttx} we present an efficient design for computing baseline stakes that meet the requirements. First, we sort all transactions in the mempool in a descending order, based on their transactions fee. Next, we fill the first block with the transactions from the mempool while monitoring the size of the block. Once the block size reaches the standard block size ($ \mathcal{SB}_{i} = \mathcal{B}_{s}$), the baseline stake is computed as the sum of the fee of all the transactions in the block. The procedure is repeated for the subsequent $l-1$ blocks. All of the blocks computed in the cycle are equal to the standard block in size, and their stakes are the sum of transaction fee of all the transactions in the block. The remaining transactions are then put into the last block ($\mathcal{B}_{i} = \mathcal{B}_{l}$). Once the algorithm finishes the execution, the smart contract will have a set of blocks and their corresponding baseline stakes ($\mathcal{ST}_{i} \in \mathcal{B}_{i}$). Since the baseline stake for a block is the sum of transaction fee, a miner's stake can be used to reimburse if he cheats.

\setlength{\textfloatsep}{1pt}
\begin{algorithm}[t]  
\SetAlgoLined\SetArgSty{}
\SetKwInOut{Input}{Input}  
\Input{$l, \mathcal{E}_{j},t_{sort} \leftarrow \text{Sorted }  \mathcal{T}(id,f,s)$ \\
}
\textbf{Initialize:} $size = 0,  fee = 0$\\
\For{$i = 1$ ... $l- 1$  } {
\ForEach{$ \mathcal{T}(id,f,s) \in  t_{sort}$ } {
    $\mathcal{B}_{i} \leftarrow \mathcal{B}_{i} \cup \mathcal{T}(id,f,s)$ \\
    $size = size + (s \in  \mathcal{T}(id,f,s))$  \\
     $fee = fee + (f \in  \mathcal{T}(id,f,s))$ \\
     \textbf{remove} $\mathcal{T}(id,f,s)$ \textbf{from} $t_{sort}$ \\
      \eIf{($size \leq \mathcal{B}_{s}$)}{
        \textbf{continue}
      }
      { 
        $\mathcal{ST}_{i} = fee$ \\
        $fee = 0$;  $size = 0$ \\ 
      } } 
      }

\For{$i = l $  } { 
\ForEach{$ \mathcal{T}(id,f,s) \in t_{sort}$ } {
    $\mathcal{B}_{i} \leftarrow  \mathcal{T}(id,f,s)$ \\
     $fee = fee + (f \in  \mathcal{T}(id,f,s))$ \\
     \textbf{remove} $\mathcal{T}(id,f,s)$ \textbf{from} $t_{sort}$
      \If{($t_{sort} = \emptyset$)}{
        $\mathcal{ST}_{i} = fee$ \\
      }
      } 
      }

\SetKwInput{KwData}{return} 
 \KwData{ $\mathcal{B}_{i}, \mathcal{ST}_{i} $} 
\caption{Computing Baseline Stake}
\label{algo:sorttx}
\end{algorithm}

\subsection{Block Auction} \label{sec:ba}
Once, the baseline stakes and epoch length are determined, the smart contract announces them to the network. The nodes query the baseline stakes corresponding to each block and compare them with their balance. A node is qualified for mining only if its balance exceeds the baseline stake. To extend mining opportunities to more peers, the smart contract forces peers to place one bid per block. It is possible that a node qualifies for multiple blocks as a candidate miner. However, a node can only place the bid on one block.

Since the blocks are sorted in a descending order based on the value of the transaction fees, a greedy candidate can select a block based on the reward incentive. The process of block auction is carried out in multiple phases. In the first phase, the qualified peers notify the smart contract and place their bids on each block. As a result, the smart contract obtains a list of candidate miners from the network ($c \subset n$), with a balance exceeding the baseline stake. In the next phase, the smart contract selects the final list of miners ($k \subset c$) and asks them to make a final commitment for their selected block. After all the commitments are locked, the smart contract assigns blocks to the corresponding miners, the miners verify each transaction, and receive the rewards. Below, we outline the details of each auction phase.

\begin{algorithm}[t]
\SetAlgoLined\SetArgSty{}
\SetKwInOut{Input}{Input}  
\Input{$\mathcal{ST}_{i}, \mathcal{B}_{i}$}  
\textbf{Initialize:} $pBlock$ \tcp*[l]{block list} 

\For{$m \in N(m_{n},b_{n}) $}{
\ForEach{ $i = 1 ... l$}{
\If{$b_{n} > \mathcal{ST}_{i}$}{ 
$pBlock \leftarrow \mathcal{B}_{i} $ \tcp*[l]{potential block} 
}}
\SetKwInput{KwData}{Output} 
 \KwData{$pBlock$ }
\textbf{Select:} $\mathcal{B}_{i} \in pBlock$  \tcp*[l]{select block} 
}
\For{$\mathcal{ST}_{i} \in \mathcal{B}_{i}$}{
$rBalance = b_{n} - \mathcal{ST}_{i}$   \tcp*[l]{balance left} 
$\mathcal{S}_{c} = \frac{(x) \times rBalance}{100}$ \tcp*[l]{stake committed} 
}
$\mathcal{P}_{c} \leftarrow \mathcal{S}_{c}$; \\

\SetKwInput{KwData}{return} 
 \KwData{$C(m_{c},b_{c}, \mathcal{P}_{c} ) $}  
\caption{Selecting Candidate Miners}
  \label{algo:sk}
\end{algorithm}

\subsubsection{Selecting Candidate Miners} \label{sec:cm}
In the first phase, the peers compare their balance with the baseline stakes of all blocks published by the smart contract. The peers then select a block of their choice for which their balance exceeds the baseline stake, and place a stake bid. In \autoref{algo:sk}, we show how an individual peer surveys the network conditions and places his bid on a certain block. The bid commitment consists of a percentage balance remaining after the baseline stake is subtracted from the current balance (lines 8-9). We take this approach for two reasons. First, the balance committed beyond the baseline stake can be later used to additionally reimburse the victim, in case the miner misbehaves. This also increases the fraud penalty for malicious miners and enforces fair mining practices. Second, the percentage commitment is useful in determining the risk factor associated with the commitment of the miner.  If a candidate miner $x$ has a higher balance $b_{x}$ than a competing miner $y$ with balance $b_{y}$, and $x$ makes 15\% commitment of ($b_{x} - \mathcal{ST}_{i}$), while $y$ makes 20\%  commitment of ($b_{y}-\mathcal{ST}_{i}$), then the preference will be given to $y$, for showing higher commitment. This method also helps in democratizing the network since the percentage stake are always in the range [0--100], where all miners will have equal opportunities to participate in the block auction. This also challenges the hegemony of rich stakeholders and decentralizes the network by offering mining opportunities to more peers.

\subsubsection{Finalizing Miners} \label{sec:fm}
Once the smart contract receives a percentage of stake committed by the candidate miners, it then finalizes the subset of miners ($ k \subset c$)  that are designated to process each block. For each block, it sorts the percentage bids committed by all candidate miners, and assigns the blocks based on the winning bids. The winning bid for each block is the highest percentage stake $\mathcal{P}_{c}$ committed by the candidate miner. However, there are cases that may cause conflicts in the final selection process. In the following, we present those cases and remedies to address them.

\BfPara{Case 1: Equal bids} It is possible to have more than one high but equal percentage stakes committed by the candidate miners. For example, a candidate miner $x$ commits 30\% of his balance for a block $\mathcal{B}_{i}$, while another candidate miner $y$ also commits 30\% of his balance for the same block. Since only one miner is selected for a block, the smart contract has to pick one of the two candidates. To resolve this conflict, the smart contract keeps a record $M(m_{j},h_{j})$ of all of the previous blocks computed and their respective miners. It then queries that record and compares how many blocks $x$ and $y$ have computed prior to the current epoch, and the one with less prior blocks is selected (\autoref{algo:sv}).

\begin{algorithm}[t]  
\SetAlgoLined\SetArgSty{}
\SetKwInput{KwData}{Input} 
 \KwData{$C(m_{c},b_{c}, \mathcal{P}_{c} ), M(m_{j},h_{j}) $ } 
\textbf{Initialize:} $cList$\\ 

\ForEach{$\mathcal{B}_{i}$}{

\eIf{$\mathcal{P}$ \text{for first item in Sorted} $C(m_{c},b_{c}, \mathcal{P}_{c})$ \text{is unique}}{
\textbf{Assign:} \text{Block to the miner at first index} \\
}

{
\text{Put all} $C(m_{c},b_{c}, \mathcal{P}_{c})$ \text{with same} $\mathcal{P}_{c}$ \text{in list}
$cList$ \\
}
\textbf{Query:} $(M(m_{j},h_{j}) )$\\
\text{sort} $cList$ {\text{based on value of} $m_{c}$ \text{in}$(M(m_{j},h_{j}) )$}\\
\textbf{Select} $m_{c}$ \text{with minimum value of} $h_{j}$ \\

\textbf{Cold Start} \text{if} $h_{j}$ \text{= 0, select one with highest bid}  \\
\textbf{Assign:} \text{Block to the Selected Miner} \\
}

\caption{Final Selection of Miners}
\label{algo:sv}
\end{algorithm}

\BfPara{Case 2: Cold start} Assuming a case where both $x$ and $y$ have not computed any block before, the contract checks the balance of both candidate miners and selects the one with a greater balance. This fulfills the third objective defined in \textsection\ref{sec:bs}, whereby miners with the higher balance benefit for their overall stake.  In \autoref{algo:sv}, we describe how the smart contract selects the final list of miners, resolves conflicts, and assigns blocks to the selected miners. In \cref{fig:abs}, we provide the abstraction of the network's state after the smart contract finalizes the list of miners for each block. 

\subsection{Block Mining} \label{sec:bm}
Once miners are finalized, the smart contract initiates a request for the final stake commitment. The final commitment ensures that a candidate miner has not spent his balance after making the initial commitment. In response, the selected miners confirm their final bids and deposit their balance to the smart contract. The smart contract locks their balance and assigns respective blocks to each miner. The miners consult the blockchain and the ``UTXO'' set to validate the authenticity of each transaction. If a transaction fraudulent, the miner notifies the smart contract and removes the transaction from the block. 

When all transactions are verified, the smart contract employs a signature scheme to prevent blockchain forks and assert ownership of the blocks. The signature scheme used by the smart contract is a tuple $(\keygen, \sign, \verify)$, where each algorithm is defined as follows. 
\begin{enumerate*}[label=\textbf{\arabic*)},start=1]
\item $\keygen(1^u)$: given a security parameter $u \in \mathbb{U}$, the key generation algorithm generates a pair of private and public keys, namely $sk$ $pk$. 
\item $\sign(sk, m)$: a deterministic algorithm that takes as an input a message $m\in\mathcal{M}$ and the private key $sk$, and generates a signature $\sigma$ corresponding to $m$. The message $m$ here is the block produced by the smart contract. 
\item $\verify(pk, m, \sigma)$: a verification algorithm that takes the public key $pk$, the message $m$, and the signature $\sigma$ as inputs, and outputs 0 if $\sigma$ is an invalid signature of $m$ and $1$ otherwise. This algorithm is used by network peers to verify the block authenticity. 
\end{enumerate*}

\begin{figure}[t]
	\centering
	\includegraphics[width=0.9\linewidth]{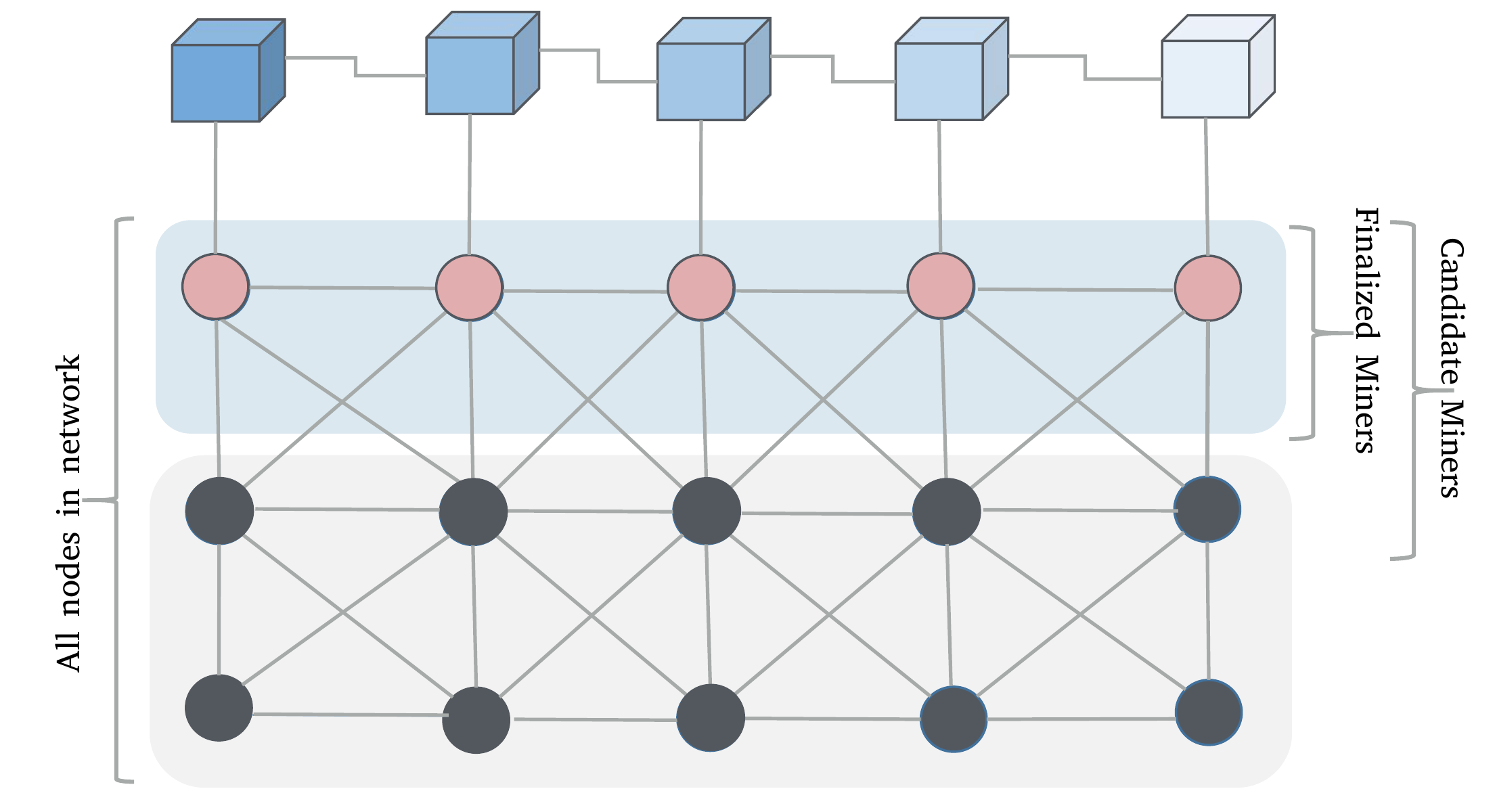}
	\caption{A overview of \ep-based blockchain system consisting of  network peers, candidate miners, and finalized miners. The shades of blocks represent their stakes and reward in descending order.   }
	\label{fig:abs}
\end{figure}

\begin{figure}
\begin{center}
    \includegraphics[width=0.45\textwidth]{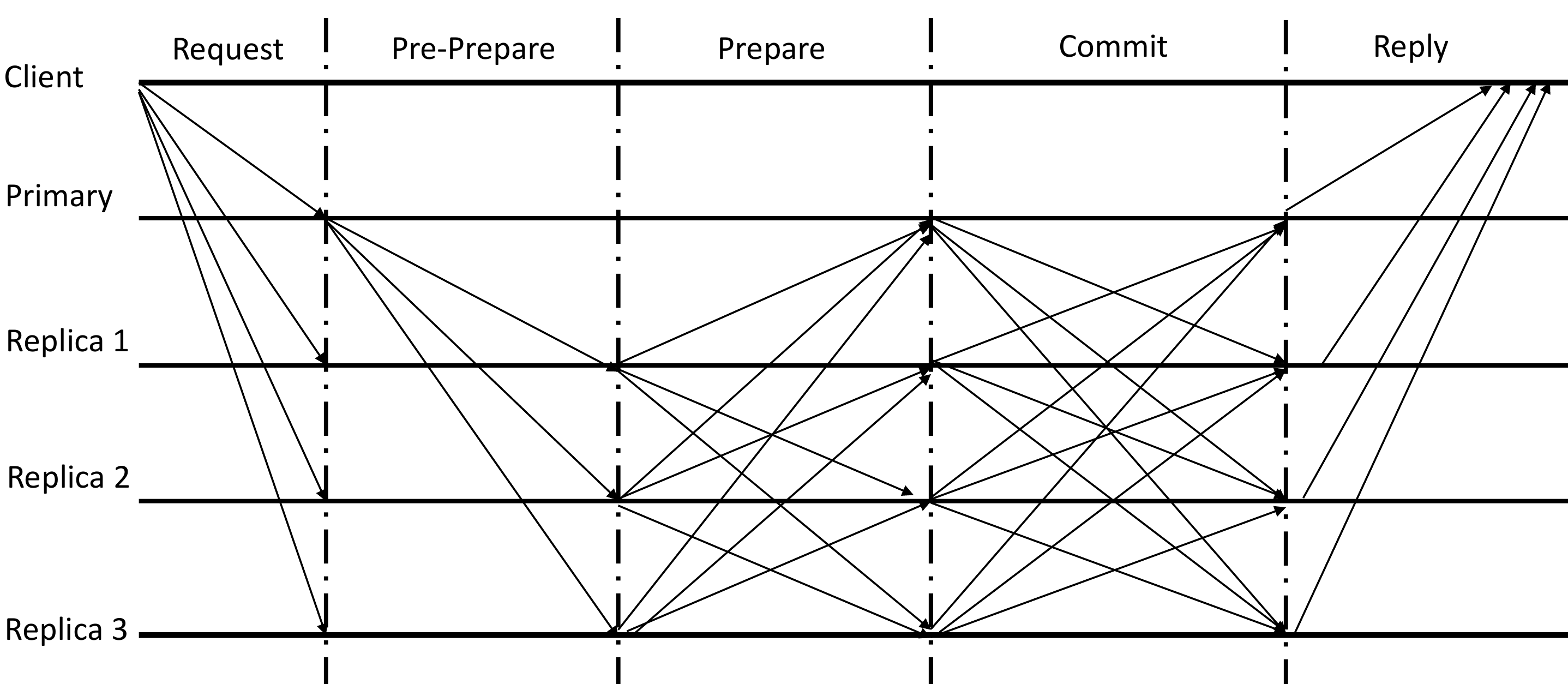}
\caption{PBFT protocol overview where client issues a request to the primary replica. The primary broadcasts the transaction to all the other replicas. The replicas validate the transaction and share their view with each other. The process of transaction verification follows four stages, namely Pre-Prepare, Prepare, Commit, and Reply. The transaction, in this case will be the mempool view issued by the primary itself.}
\label{fig:PBFT}
\end{center}
\end{figure}

\begin{figure*}
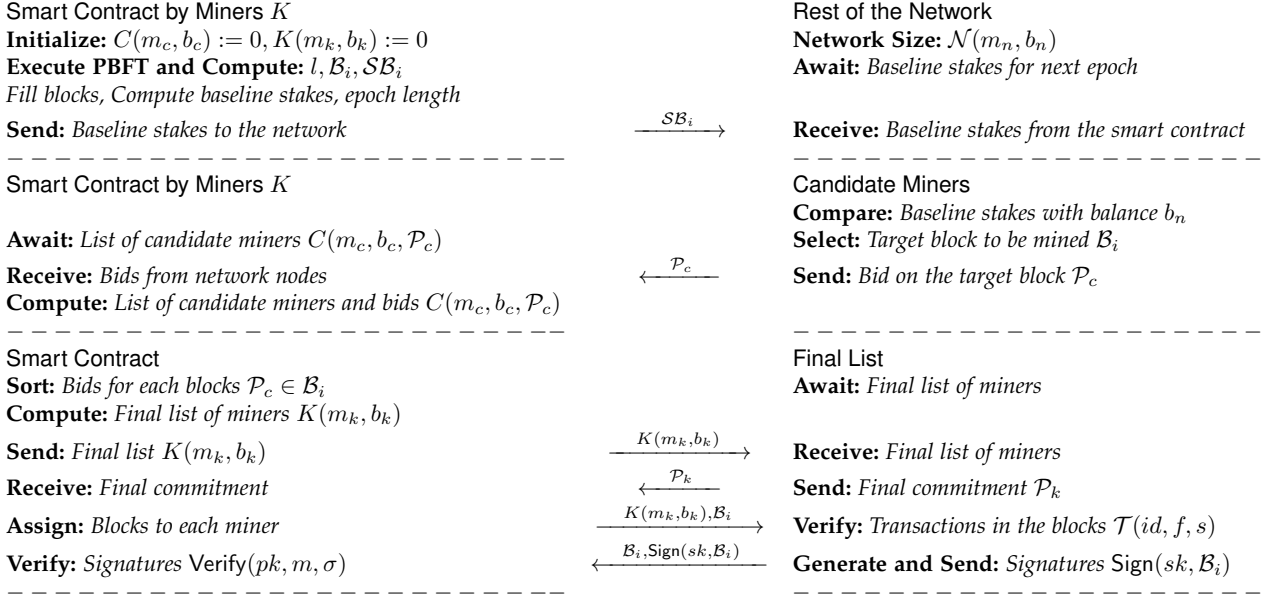


  \begin{adjustbox}{minipage=\linewidth,scale = 0.9 }
\[
\begin{array}{l c l}
\text{\textsf{Smart Contract by Miners $K$}} & & \text{\textsf{Rest of the Network}} \\
\textbf{Initialize: } C(m_{c},b_{c}) := 0,  K(m_{k},b_{k}) := 0 & &  \textbf{Network Size: } \mathcal{N}(m_{n},b_{n})\\ 

\textbf{Execute PBFT and Compute: } l, \mathcal{B}_{i}, \mathcal{SB}_{i} & & \textbf{Await: } \textit{Baseline stakes for next epoch} \\
\textit{Fill blocks, Compute baseline stakes, epoch length} & & \\
\textbf{Send: } \textit{Baseline stakes to the network}& \xrightarrow{\hspace{1em} \mathcal{SB}_{i}\hspace{1em}} & \textbf{Receive: } \textit{Baseline stakes from the smart contract}\\

------------------------ & & ----------------------- \\

\text{\textsf{Smart Contract by Miners $K$}} & & \text{\textsf{Candidate Miners}} \\
 & & \textbf{Compare: } \textit{Baseline stakes with balance $b_{n}$} \\
\textbf{Await: } \textit{List of candidate miners } C(m_{c},b_{c}, \mathcal{P}_{c} ) & & \textbf{Select: } \textit{Target block to be mined } \mathcal{B}_{i} \\
\textbf{Receive: } \textit{Bids from network nodes} & \xleftarrow{\hspace{1em} \mathcal{P}_{c}\hspace{1em}} & \textbf{Send: } \textit{Bid on the target block } \mathcal{P}_{c} \\ 
\textbf{Compute: } \textit{List of candidate miners and bids } C(m_{c},b_{c}, \mathcal{P}_{c} ) & &  \\
------------------------ & & ----------------------- \\

\text{\textsf{Smart Contract}} & & \text{\textsf{Final List}} \\
\textbf{Sort: } \textit{Bids for each blocks } \mathcal{P}_{c} \in \mathcal{B}_{i} & & \textbf{Await: } \textit{Final list of miners} \\
\textbf{Compute: }  \textit{Final list of miners }K(m_{k},b_{k}) & & \\
\textbf{Send: } \textit{Final list } K(m_{k},b_{k}) & \xrightarrow{\hspace{1em} K(m_{k},b_{k})\hspace{1em}} &  \textbf{Receive: } \textit{Final list of miners} \\
\textbf{Receive: } \textit{Final commitment} &   \xleftarrow{\hspace{1em} \mathcal{P}_{k}\hspace{1em}}& \textbf{Send: } \textit{Final commitment }  \mathcal{P}_{k}  \\ 
\textbf{Assign: } \textit{Blocks to each miner} &   \xrightarrow{\hspace{1em} K(m_{k},b_{k}), \mathcal{B}_{i}\hspace{1em}}&  \textbf{Verify: } \textit{Transactions in the blocks }  \mathcal{T}(id,f,s) \\
\textbf{Verify: } \textit{Signatures } \verify(pk, m, \sigma) &  \xleftarrow{\hspace{1em}  \mathcal{B}_{i}, \sign(sk, \mathcal{B}_{i})\hspace{1em}} & \textbf{Generate and Send: } \textit{Signatures }  \sign(sk, \mathcal{B}_{i})   \\

------------------------ & & ----------------------- \\

\end{array}
\]
  \end{adjustbox}

\caption{Information workflow between smart contract executed by $K$ for the next epoch. The smart contract computes baseline stakes and receives bids for each block. It then selects the miner list $K$ for the next round. Rewards are released only when the next epoch ends successfully. }
\label{fig:protocol}
\end{figure*}

Once the miner verifies all transactions, he generates a public/private key pair, and signs the message  $\mathcal{B}_{i}$, with his private key  $\sign(sk, \mathcal{B}_{i})$, computes the hash of the block $H(\mathcal{B}_{i})$, and sends the block to the smart contract. The smart contract authenticates the block by using the public key $pk$ of the miner and releases the block to the network. Upon receiving the block, nodes in the network also validate the miner's signatures. This process prevents forks in case an adversary publishes a counterfeit block to the network. 

When an epoch ends after $\mathcal{D}_{ep}$, the mempools at each node are filled by transactions exchanged during the epoch. The probability of $z$ number of transactions arriving at a mempool during the epoch can be computed using \autoref{eq5}, and the new size of mempool $\mathcal{M}_{s}$ can be calculated using \autoref{eq6}.  
\begin{align} \label{eq5}
 P (z \text{ arrivals in } T) &= \frac{(\lambda T)^{z} e^{\lambda T} }{z\,!} = \frac{(\lambda \mathcal{D}_{ep})^{z} e^{\lambda \mathcal{D}_{ep}} }{z\,!} \\
\label{eq6} \mathcal{M}_{s} = \lambda \times T &=  \lambda \times \mathcal{B}_{i} \times \mathcal{B}_{t} = \lambda \times \mathcal{D}_{ep}
\end{align}

In the next step, the miners $k\subset c$ calculate the parameters for the next epoch. For parameter calculation, miners must agree upon the mempool state so that their baseline stakes, block size, and epoch lengths are consistent. In blockchain systems, mempool states can vary across nodes. In Bitcoin, for example, the mempool state can change due to the varying mempool size limits specified by each node or the network asynchrony. In Bitcoin, nodes can arbitrarily set the mempool size to prevent RAM overloading. Therefore, nodes with different RAM sizes have different mempool sizes, which can cause inconsistency in our model. 

The mempool size can also vary due to network asynchonry typically created by (1) the transaction propagation delay, and (2) the network topology. Since Bitcoin nodes are spread across autonomous systems~\cite{SaadCLTM19,SaadAAAYM19}, the transaction exchange among nodes can experience varying delay. As a result, nodes that experience high transaction propagation delay may not receive a transaction in their mempool prior to executing \ep. Moreover, at any time, there are $\approx$10K Bitcoin nodes, and the default connectivity limit is 125~\cite{SaadCLTM19}. Therefore, the network does not form a completely connected topology due to which the transaction propagation does not follow the broadcast model~\cite{GarayKL17}. As a result, the mempool state can vary across nodes at any time. To prevent this inconsistency, $K$ must agree upon a common mempool state to output consistent parameters for the next epoch.

For a common the mempool state, the miners execute PBFT protocol. In \cref{fig:PBFT}, we give an overview of the PBFT-based consensus and for more details we refer the reader to \cite{Castro00}. In a \cc network, where the number of peers can be in the thousands, applying PBFT can be infeasible. However, in \ep, PBFT is executed among miners. Since $k$ in an epoch is much less than $n$, PBFT can be applied efficiently. The smart contract randomly specifies a primary replica in $k$ which will share its mempool view with other replicas and obtain consensus. To ensure that all replicas faithfully execute the protocol, we create a dependency between epochs. Miners' rewards are not released by the contract until the next epoch is successfully initiated. Therefore, to favor self interest, miners will initiate the next epoch and its smart contract. In~\cref{fig:protocol}, we provide a complete workflow of \ep starting from computing the baseline stake to block mining by the final list of miners $K$.

\BfPara{Benefits and Limitation of Using Mempool} In \ep we use the mempool to derive the epoch length and the baseline stakes. Our mempool-based approach is novel since it has not be adopted by existing PoS models. In this section, we succinctly summarize the key aspects of our technique.  

In blockchain systems, mempool is a repository of unconfirmed transactions from which miners select transactions for blocks. Ideally, if the block size is equal to the mempool size, there is no transaction backlog and all transactions are processed within the expected time. However, in practice, there is usually a transaction backlog that can harm the blockchain system as shown in~\cite{SaadTM18}.

Acknowledging the role of a mempool in transaction processing, we use it to determine the epoch length. Our approach is indeed innovative since the existing PoS-based schemes do not use mempool to determine the epoch length. Moreover, this approach gives us two additional benefits. The first benefit is that based on the epoch length and duration, \ep can guarantee a new block at the average block time. In the existing blockchain system (\ie Bitcoin), while the expected time for a new block is $\approx$10 minutes, the inter-arrival time between two blocks cannot be guaranteed to follow the expected time. As a result, the transaction confirmation can be delayed. In \ep, we address this problem by fixing the epoch duration. 

The second benefit is that mempool allows us to determine the baseline stake for each block. In the existing blockchain systems, miners tend to prioritize transactions based on the transaction fee. Similarly, in \ep, we sort transactions based on the transaction fee in different blocks. Each block has a different transaction fee and a candidate miner with a higher balance can bid on a block with a higher fee. This bidding process ensures that miners with a higher balance are rewarded proportional to the stake that they commit. If any miner misbehaves during the protocol, his stakes are used to compensate the affected parties.

While the mempool-based approach has several benefits, it has a limitation as well. As mentioned earlier, the mempool state can vary across miners due to which they may compute different baseline stakes and epoch length. To prevent that, we use PBFT to ensure that miners agree upon consistent values of baseline stakes and epoch lengths.

\section{Feasibility Analysis of \ep} \label{sec:analysis}

\subsection{Meeting Requirements} 
\BfPara{Decentralization} To achieve decentralization and overcome skews associated with the conventional PoS, \ep applies two conditions: percentage stake commitment and one block per miner approach during an epoch. The percentage stake provides equal opportunities to miners to place a bid on the block of their choice, thereby increasing decentralization. One block per epoch ensures that an almost unique set of miners benefits from the fee rewards during the epoch. This also contributes to decentralization and encourages fair mining. Another feature of \ep is the conflict resolution when more than one miner meet the criteria. For instance, if all miners put 99\% balance as stake beyond the baseline stake, the smart contract consults its mining repository to select the miner who has mined less blocks in the past.

\BfPara{Relative Reward} Another objective is the relative reward of miners with respect to their stakes. If a miner puts higher stakes than others, then he must also receive higher mining rewards. We meet that condition by sorting transactions based on their mining fee and calculating baseline stakes accordingly. In \ep, the first block $\mathcal{B}_{1}$ and its corresponding baseline stake $\mathcal{SB}_{1}$ are greater than the subsequent blocks and their baseline stakes. Furthermore, the last block $\mathcal{B}_{l}$ and its baseline stake $\mathcal{SB}_{l}$ are less in size and reward than any other block in the epoch. Therefore, a rich miner can always opt for the blocks higher in the hierarchy of the chain to obtain more rewards, while a poor miner can select blocks from a lower hierarchy of the chain. This scheme is also applied during a conflict between two miners who put equal stakes for a target block. If the two miners have no history in the mining repository of the smart contract, a miner with higher balance is selected. Therefore, \ep also considers each miner based on his risk potential. In contrast to the other notable schemes~\cite{KiayiasRDO17,ChenM19,DaianPS19}, where miners are selected at random, in \ep, we compliment each miner based on the relative stake that they commit.

\subsection{Security Analysis of \ep} \label{sec:sa}
In this section, we evaluate \ep under notable PoS-based attacks. Previously, in \autoref{sec:bm}, we showed that the signature scheme prevents blockchain forks. Some other attacks discussed in this section are the majority attacks, stake theft, and nothing-at-stake. In the following, we analyze each attack in an \ep-based system. For each attack, we assume a polynomial-time adversary $\mathcal{A}$, with knowledge of \ep. For the majority attacks, we assume the $\mathcal{A}$ has accumulated more than 50\% stake in the system. Although the majority attacks are naturally less probable in PoS compared to PoW, since accumulating 50\% stakes in a system is more difficult than acquiring 50\% hash rate. However, assuming the worst case, we will demonstrate that \ep makes the attack infeasible for an adversary with 50\% stakes. For stake theft and nothing-at-stake, we assume the adversary knows about the operations and communication model of \ep, and the adversary tries to cheat or halt the auction process.   

\BfPara{Majority Attack} First, we analyze the notion of fairness defined in \tsref{sec:fairness}, in which the adversary with 50\% stakes (1) launches a majority attack to double-spend, and (2) wins each subsequent block auction. In \ep, we show that the probability successfully launching a majority attack is negligible~\autoref{eq:adv2}, and the probability of mining consecutive blocks with 50\% hash rate is less than 1~\autoref{eq:nadv3}. 

To show that \ep achieves fairness, assume $n$ peers in the system with a combined balance of $\alpha+\beta$ coins. Further assume that the adversary acquires ($\alpha$=0.51) coins, and $\beta$ is uniformly divided among the remaining $n-1$ peers. In the conventional design, \autoref{eq:nadv2} will be 1 and the adversary is guaranteed to win the next block. In comparison, assume that all peers in \ep are greedy, making equal bids of 99\% of their balance. In that case, the probability that the adversary mines the next block becomes:
\begin{align}
\label{eq:nadv5}\text{\normalfont Pr} (~\mathcal{A}  \text{\normalfont ~mines the next block in \ep} ~|~\alpha\geq 0.5) = \frac{1}{n}
\end{align}

This is to be noted that if the adversary splits his stake among $p$ nodes and each node places a bid of 99\%, then the success probability can increase from $1/n$ to $p/n$. Nevertheless, as long as there is one honest party in the system, $p<n$, the probability remains less than 1. In our system, $p=n$ means that the adversary controls all nodes in the system which is an infeasible assumption in practice. Another way to express this is that as long as the adversary does not hold 100\% coins in the system, his probability of winning the next block auction remains less than 1. In contrast, in the conventional PoS models, with 51\% coins, the adversary wins the next auction with probability 1~\autoref{eq:nadv2}.

Next, we analyze the success probability of launching a majority attack (\ie mining a longer private chain with $m$ consecutive blocks) to replace the public chain. Note that in \ep, the smart contract maintains a record of miners who mine blocks. Provided that, we present two cases in which the adversary launches a majority attack. In the first case, we assume the adversary does not control any other node in the system. In the second case, we assume the adversary controls $p$ nodes in the system, divides his balance among them, and allows them to put a maximum bid on each block. For the first case, the probability that the adversary mines two successive blocks with $\alpha\geq0.5$ becomes $(\frac{1}{n})\times (\frac{1}{n-1})$. Therefore, the probability that adversary mines $m$ subsequent blocks to double-spend becomes: 
\begin{align}
\label{eq:adv7}\text{\normalfont Pr} (~\mathcal{A}  \text{\normalfont ~double-spends in \ep} ~|~\alpha\geq 0.5)  = \prod_{i=0}^{m} \frac{1}{n-i}. 
\end{align}
Eq.~\autoref{eq:adv7} shows that for a large value of $n$ (a large network size), the probability of double-spending through a majority attack becomes negligible. For the second case, the adversary's probability of mining the first block becomes $p/n$, and the probability of mining the subsequent block becomes $(p-1)/(n-1)$. As such, the probability that adversary mines $m$ subsequent blocks to double-spend becomes: 
\begin{align}
\label{eq:nadv7}\text{\normalfont Pr} (~\mathcal{A}  \text{\normalfont ~double-spends in \ep} ~|~\alpha\geq 0.5)  = \prod_{i=0}^{m} \frac{p-i}{n-i}.
\end{align}
In Eq.~\autoref{eq:nadv7}, keeping a realistic assumption that $p\neq n$, the adversary cannot successfully double-spend with probability 1. In other PoS models, with 51\% coins, the adversary could double-spend with probability 1~\autoref{eq:adv2}. Our analysis shows that \ep achieves more fairness than conventional PoS models. 

\BfPara{Stake Theft} If the adversary is a malicious candidate miner who commits a very high stake $\mathcal{P}_{c}$ to the smart contract, and later spends it, he might be able to game the system. However, to counter that, we introduced the final stake commitment in \textsection\ref{sec:bm}. The smart contract confirms that the final stake commitment is equal to the initial commitment to proceed with the block assignment. Therefore, it prevents a malicious candidate miner from gaming the system.

If the adversary is among the selected miners, and invalidates one or more transactions in a block, the smart contract will punish the adversary by awarding his stake to the victim. Since the baseline stake for each block ($\mathcal{ST}_{i} \in \mathcal{B}_{i} $) is the cumulative sum of transaction fee in the block, the baseline stake acts as an insurance. If one or more transactions are compromised, the baseline stake covers them by reimbursing their fee. The penalty is derived from the final stake committed by the selected miner. Since the final stake is the percentage balance remaining after the baseline stake, therefore, the difference ($\mathcal{P}_{c} - \mathcal{ST}_{i}$) is used as a penalty.

\BfPara{Nothing at Stake}
An argument against the $\textsl{non-blocking}$ progress of PoS-based applications is called $\textsl{nothing-at-stake}$ \cite{martinez18}. The argument states that a PoS application would halt if no miner puts anything on stake. In \ep, this condition is less likely to arise due to following reasons. First, the problem of nothing-at-stake occurs due to centralization of mining. If there are fewer miners in the system, they can collude to halt the progress. In \ep, our main goal is to break that collusion and centralization. As such, and since mining opportunity is offered to a wider set of stakeholders (see \autoref{tab:data}), \ep will provide equal or better $\textsl{non-blocking}$ progress than other PoS schemes. Additionally, the issuer or recipient of a transaction would naturally want a non-blocking progress of the system so that their transactions are confirmed in time. As such, if there is no miner participating in the auction process, the smart contract can simply lower the threshold for the baseline stake to allow the sender or the receiver of the transaction to mine the block.

\BfPara{PBFT Fault Tolerance} As mentioned in \autoref{sec:bm}, $k$ miners in an epoch execute PBFT to obtain a common mempool view. In PBFT, assuming $f$ faulty replicas, the system requires $3f+1$ replicas to behave honestly \cite{Castro00}. In other words, $\approx$ 70\% miners are expected to behave honestly for successful execution of \ep. To motivate honest mining, we create a dependency among epochs such that the smart contract does not release rewards to miners until the next epoch is executed correctly. Moreover, if any faulty replica is detected, its rewards can be withheld as a penalty. 

\begin{figure*}

\hfill
\begin{subfigure}[Random selection Epoch length = 200 \label{fig:rs50}]{\includegraphics[width=5.5cm]{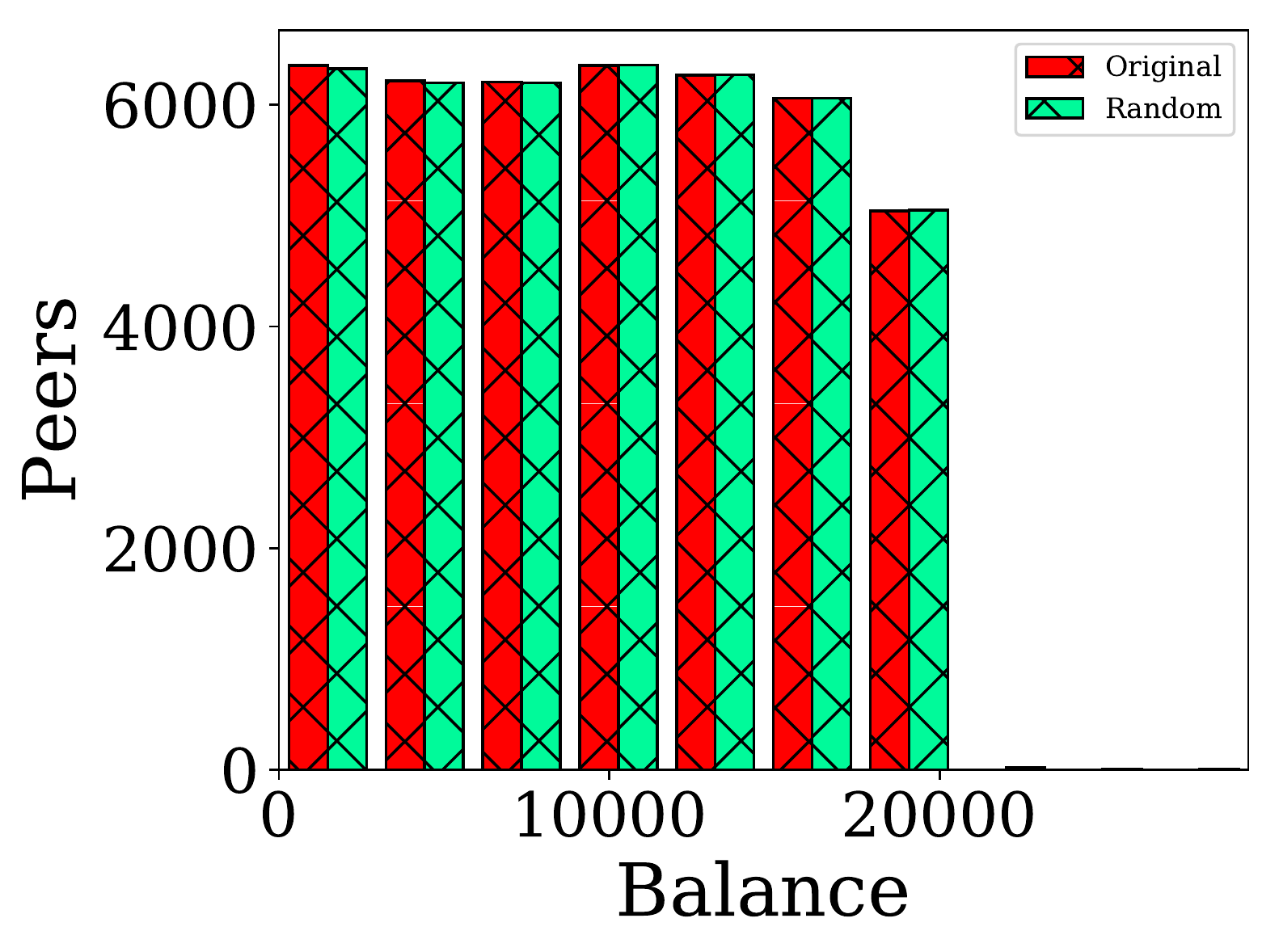}} 
\hfill
\end{subfigure}
\begin{subfigure}[Priority selection Epoch length = 200 \label{fig:rs100}]{\includegraphics[width=5.5cm]{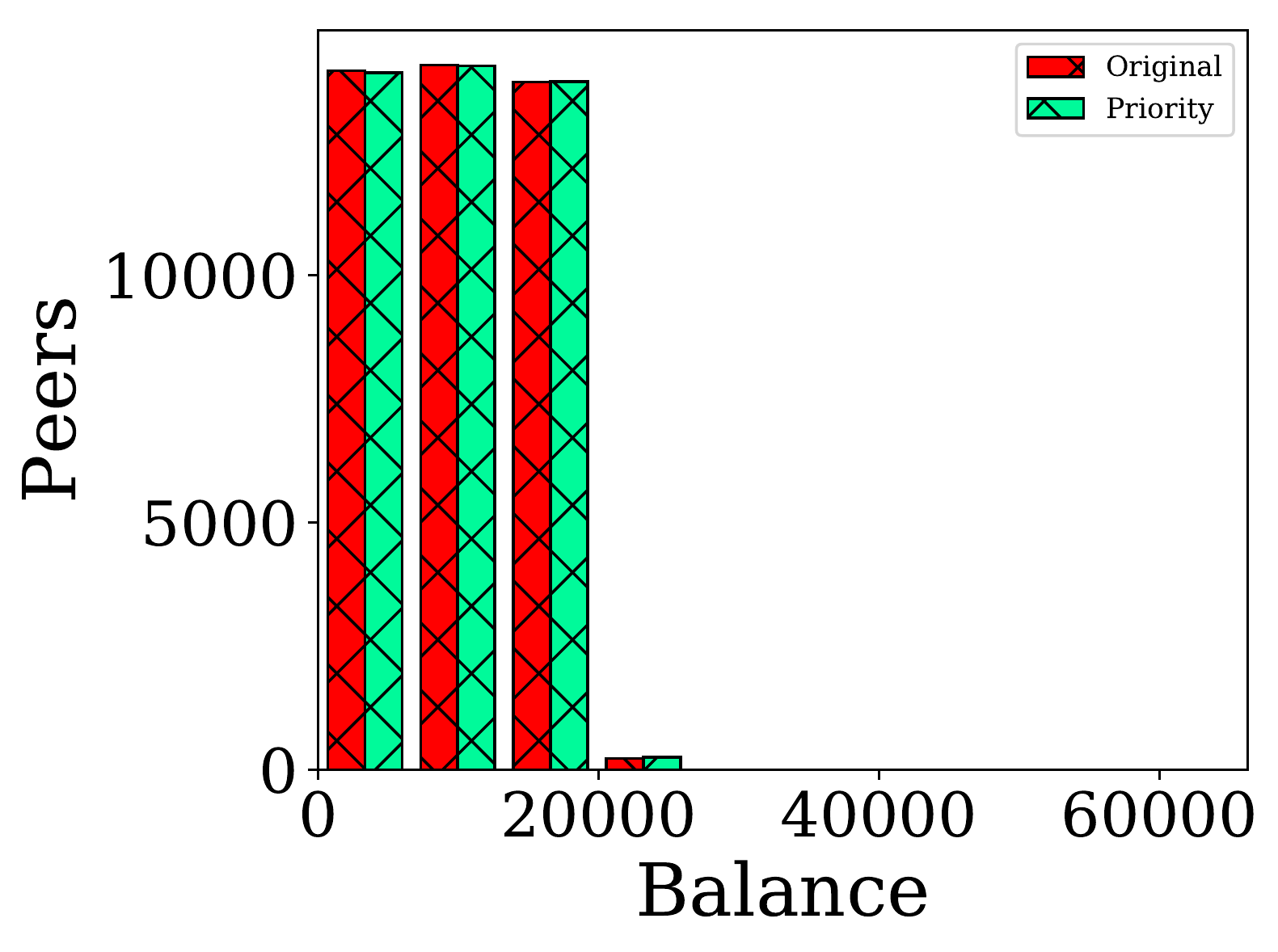}}
\hfill
\end{subfigure}
\begin{subfigure}[\ep Epoch length = 200 \label{fig:rs200}]{\includegraphics[width=5.5cm]{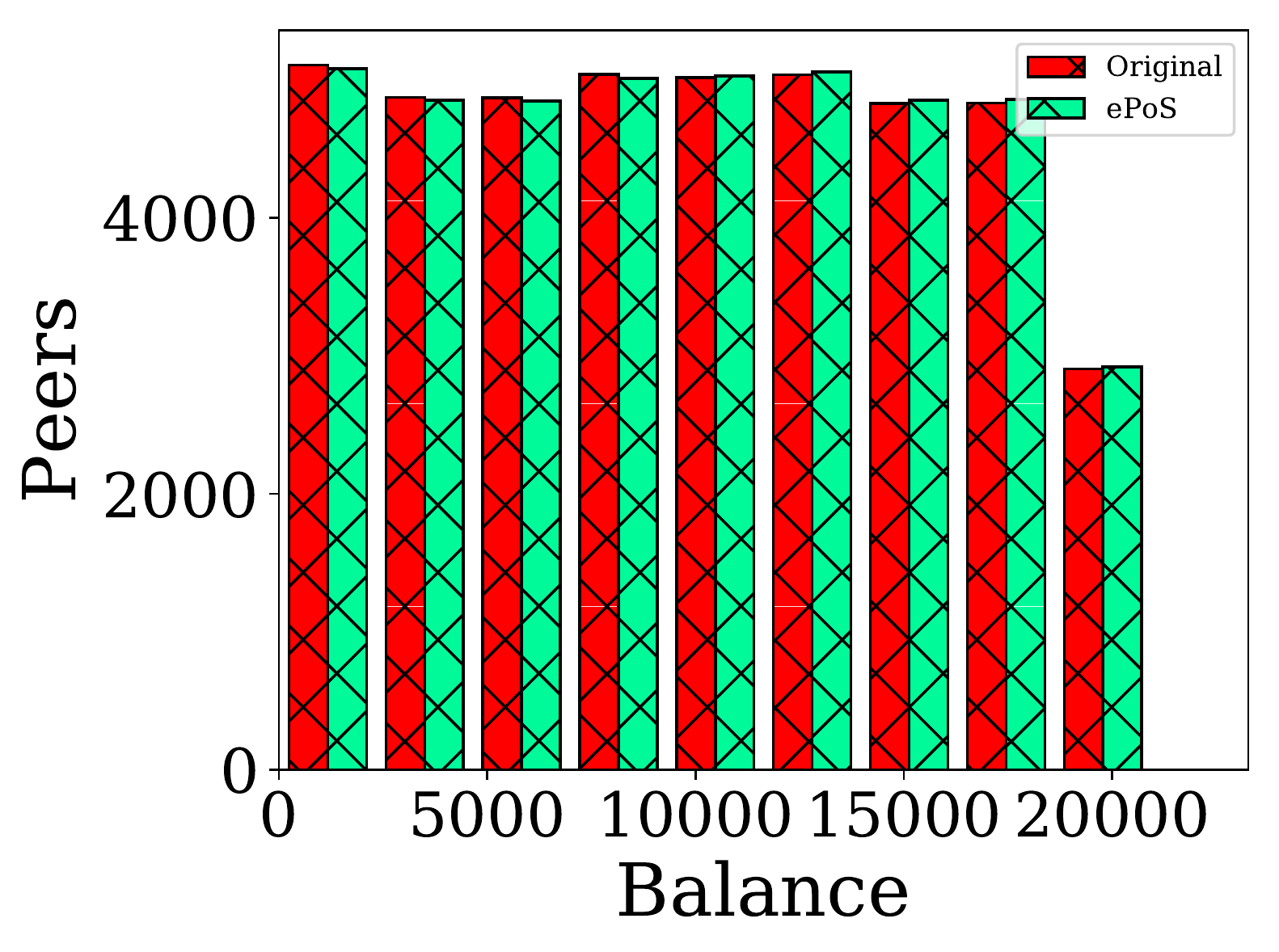}} 
\end{subfigure}\vspace{-3mm}
\caption{Simulation results. Original is the balance distribution prior to the execution. Accordingly, Random, Priority, and \ep are the balances after executing the three schemes. The wider gap on the x-axis means that the overall stake in the system has skewed after the execution. } 

\label{fig:LSTMP}\vspace{-3mm}
\end{figure*}

\begin{figure*}[!t]
\hfill
\begin{subfigure}[Random Selection
\label{fig:rs}]{\includegraphics[width=0.3\textwidth]{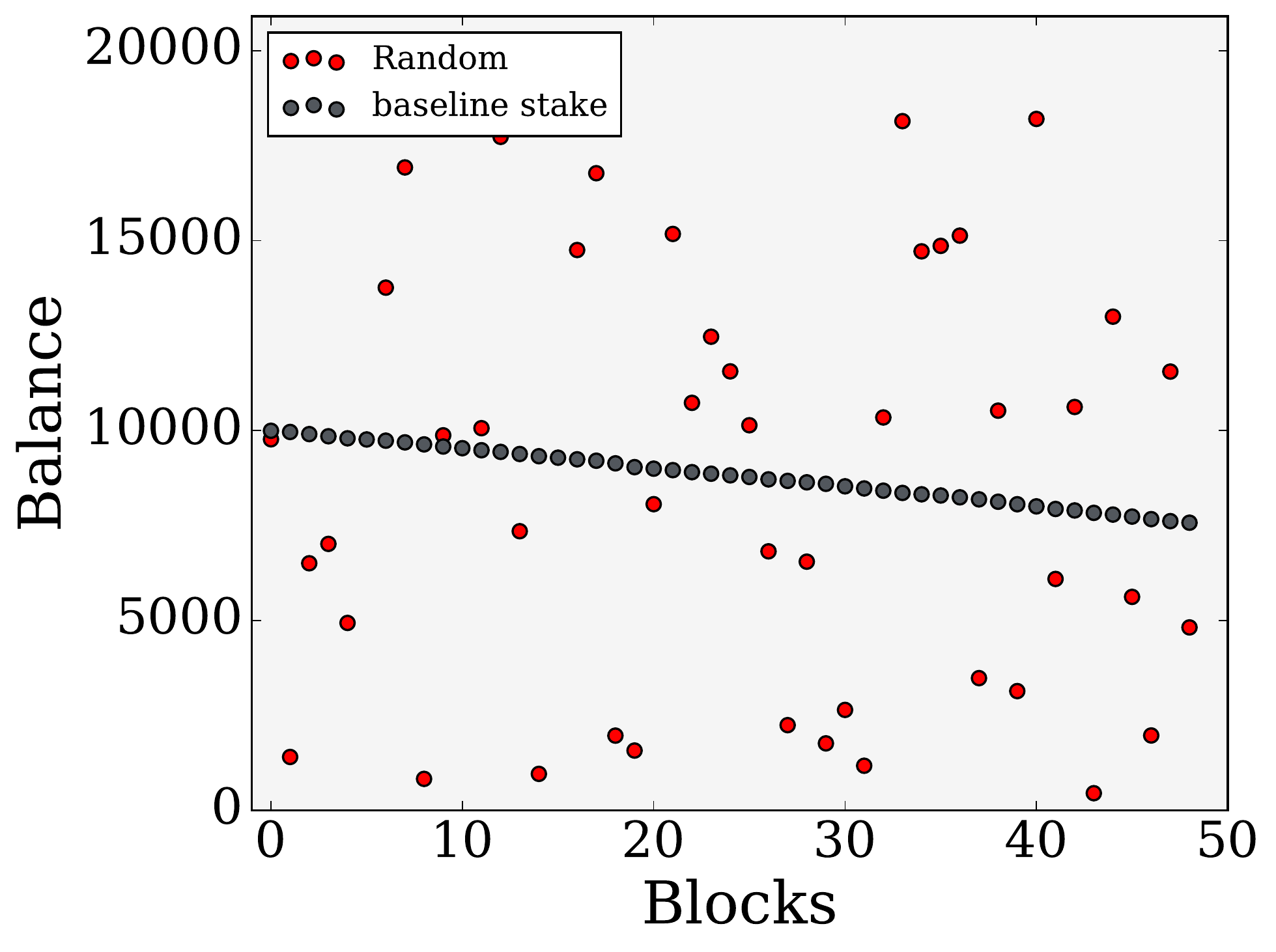}} 
\hfill
\end{subfigure}
\begin{subfigure}[Priority Selection \label{fig:ps}]{\includegraphics[width=0.3\textwidth]{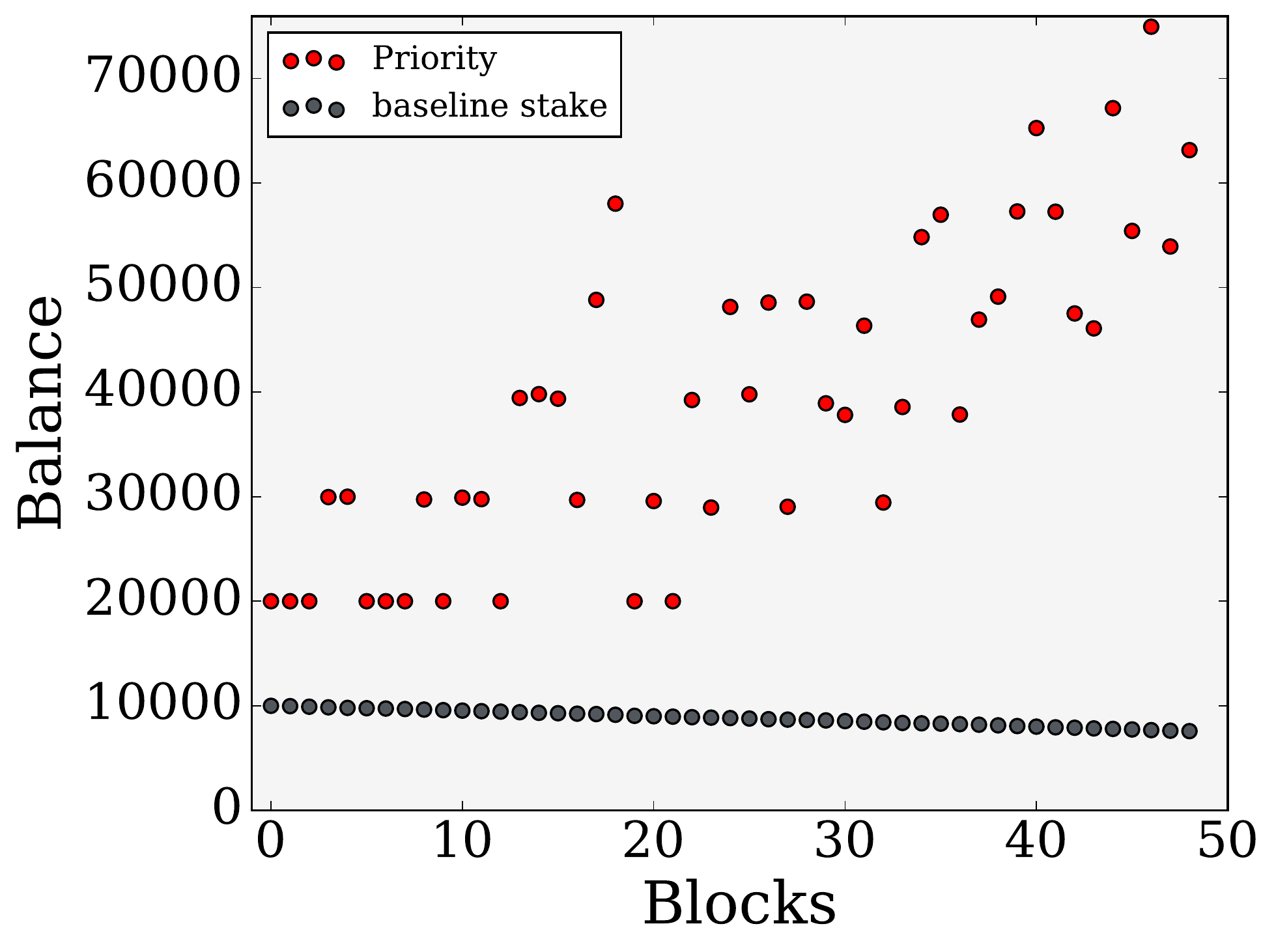}}
\hfill
\end{subfigure}
\begin{subfigure}[\ep Selection \label{fig:ep}]{\includegraphics[width=0.3\textwidth]{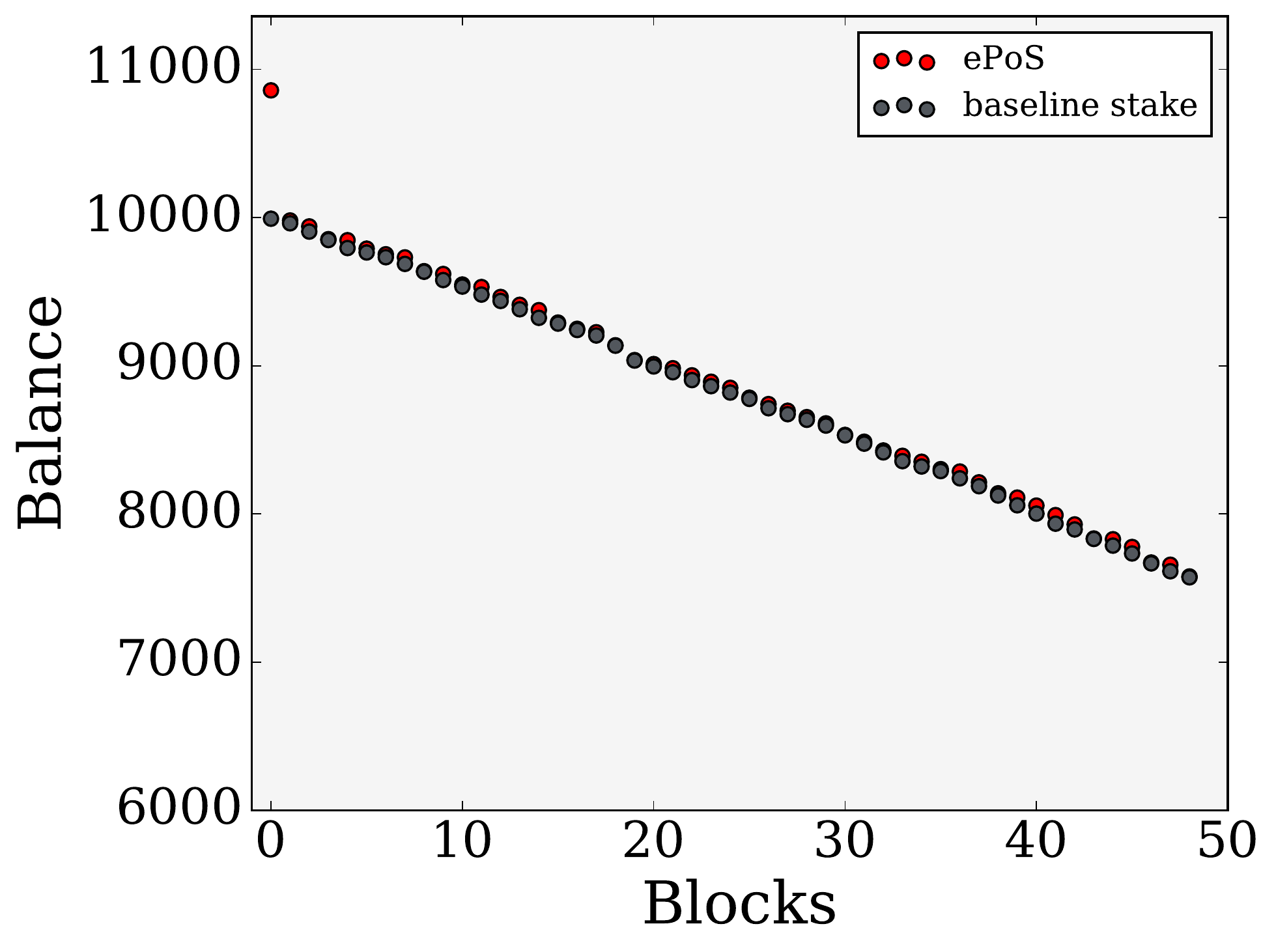}} 
\hfill
\end{subfigure}\vspace{-3mm}
\caption{Comparison of baseline stake $\mathcal{ST}_{i}$ and the balance in  the three scheme. In the random selection, half of the miners have balance less than $\mathcal{ST}_{i} \in \mathcal{B}_{i}$. This scheme is unfair since miners can easily cheat. In \ep, the balances are marginally greater than the baseline stake.  }\vspace{-5mm}
\label{fig:bsc}
\end{figure*}

\section{Simulations and Results} \label{sec:sm}
Now we present the simulations results to validate the performance of \ep compared to conventional PoS schemes. We construct three PoS models, namely the random selection, priority selection, and \ep selection. In the random selection, a miner is randomly chosen from the network to mine the next block. The miner selects the highest paying transactions from the mempool and publishes the block. In priority selection, and following the conventional PoS scheme the miner with more balance is selected for the next block. This design is similar to the existing PoS models applied in cryptocurrencies such as Peercoin, where a preference is given to the rich miners.
To capture that, we assume a subset of greedy miners in the priority selection that participate in every block auction and try to win using their high balance. Finally, in the \ep selection, we implement the extended PoS scheme as described in \textsection\ref{sec:pof}. Miners with balance greater than the baseline stake are allowed to place a percentage bid for the block auction.

\subsection{Network Overview} \label{sec:no}
To faithfully model a system closer to the existing blockchain applications, we randomly select a network size in the range of 8,000--9,000 nodes. This is close to the average size of Bitcoin network ~\cite{bitnodes_18}. For each node, we assign a balance $b_{n}$ randomly distributed between 0 and 20,000 coins. In Bitcoin, the average number of transactions per block is 2,000. To reflect that, we fix the standard block size $\mathcal{B}_{s}$ to accommodate only 2,000 transactions. Next, we randomly select the mempool size $\mathcal{M}_{s}$ between 1--100 blocks to obtain the epoch length as defined in \autoref{eq2}. For each scheme in the experiment, we run the simulation for $l$ = 200, and evaluate the balance of nodes towards the end of epoch.

\subsection{Evaluation Parameters} \label{sec:ep}
We evaluate each design based on its capability of providing decentralization and fairness guarantees. Decentralization is measured by $\beta_{e}$ and $\beta_{c}$ from \autoref{eq:decen}. Additionally, to evaluate decentralization in the case of random selection, we use $\beta_{r}$. As shown in \autoref{eq:decen}, the greater the value of $\gamma$, the more decentralized a scheme is compared to the other. Furthermore, the fairness and security is measured by comparing the balance of the miners with the baseline stake. If the balance before the auction is greater than or equal to the baseline stake ($b_{k} \in  K(m_{k},b_{k} \geq \mathcal{ST}_{i}$), the system is considered as secure. We evaluate each design under these conditions.

\subsection{Results and Evaluation} \label{sec:se}
In \cref{fig:LSTMP}, and \cref{fig:bsc}, we report the simulation results. In \cref{fig:LSTMP}, we cluster the balance of peers at the start and end of each epoch to capture the distribution of rewards among peers. Clustering the balance in histogram bins is useful in presenting the overall state of the network without showing the balance of each node. The figure shows that in priority selection, a subset of rich miners is able to win each auction and obtain all fee rewards. The accumulation of rewards among few miners creates a skew, thereby impacting the decentralization property of the scheme. As shown on the x-axis, the balance of these miners increases after the epoch, and there is little to no change in the balance of other miners. 

The random selection scheme achieves a higher decentralization by distributing block rewards among randomly selected set of peers in the system. The maximum balance of peers, shown on the x-axis, is less compared to the priority selection scheme, showing the reward distribution among a larger set of miners. The x-axis shows that the maximum balance achieved by a cluster of miners is much less (25,000) compared to the priority selection scheme (40,000). Moreover, y-axis shows the number of peers in each cluster, indicating that the lower the height of histogram on y-axis, the more distributed is the block reward.

\cref{fig:LSTMP} also shows that compared to the other two schemes \ep is more decentralized and distributes rewards to the maximum set of candidate miners. It can be observed  that the range of the x-axis, denoting the maximum balance acquired by any user, is less than the two other schemes.

To evaluate the security we plot the baseline stake of each block in the first epoch, and the respective balance of each miner. As defined in \textsection\ref{sec:ep}, if the balance of the miner $b_{k} \in K(m_{k},b_{k})$ is greater than the baseline stake $\mathcal{ST}_{i}$, the miner will be disincentivized from cheating. Our plot in \cref{fig:bsc} shows that the random selection scheme is unfair and vulnerable to attacks, since half of the miners had lower balance than the baseline stake. In comparison, the priority selection and \ep are resilient against attacks since each miner had a higher relative balance than the baseline stake. In \ep, the balance of the miner is marginally greater than the required threshold, showing a trade-off between decentralization and security, and \ep meets both  requirements.

In \cref{tab:data}, we report the empirical results obtained from the simulations. In particular, we highlight the decentralization parameter $\beta$ obtained from \autoref{eq:decen}, the unique number of miners, mean baseline stake, and the mean balance of miners prior to block mining. It can be observed that as the number of blocks per epoch increases, the decentralization parameter of \ep increases more than random selection and priority selection. More specifically, and by deriving $\gamma$ from \autoref{eq:decen}, we observe that $\beta_{e} - \beta_{c}$= 0.005, and  $\beta_{e} - \beta_{r}$= 0.001 after the third epoch. Since we asserted that if the value of $\gamma$ is greater than 0, then \ep is deemed to achieve higher decentralization than its competitive schemes. Simulation results complement our assertion.

\section{Applications of \ep} \label{sec:app}

\begin{table}[t]
\centering
  \caption{Simulation results recorded for each scheme. Mean $\mathcal{ST}_{i}$ is the average of baseline stakes of all the blocks in the epoch. Mean $b_{k}$ is the average balance of all the miners before mining. Unq $k_{i}$ are unique number of miners selected.   } \vspace{-3mm}
\scalebox{0.8}{
\begin{tabular}{|l|c|c|c|c|l|c|}
\Xhline{3\arrayrulewidth}
Scheme & $l$ & Mean $\mathcal{ST}_{i}$ & Mean $b_{k}$ & Unq $k_{i}$ & $\ \ \ \beta$  & Mean $\gamma$  \\ \Xhline{2\arrayrulewidth}
\multirow{ 3}{*}{Random} & 50                    & 8626.2                  & 10056.9                   & 50            &     0.0062 &     0     \\ \cline{2-7}
& 100                   & 7346.2                  & 9778.3                            & 96           & 0.0120  & 0.0005        \\ \cline{2-7}
& 200                   & 4892.7                  & 9834.7                        & 192           & 0.0240 & 0.0010         \\ \Xhline{2\arrayrulewidth}

\multirow{ 3}{*}{Priority} & 50                    & 8626.2                  & 53680.7                 & 10            & 0.0012   & 0.0005        \\ \cline{2-7}
& 100                   & 7346.2                  & 34614.7                     & 60            & 0.0075  & 0.005        \\ \cline{2-7}
& 200                   & 4892.7                  & 23327.9              & 160           & 0.0200  &  0.005      \\ \Xhline{2\arrayrulewidth}

\multirow{ 3}{*}{\ep} & 50                    & 8626.2                  & 8783.7                 & 50            & 0.0062      & 0   \\ \cline{2-7}
& 100                   & 7346.2                  & 7435.1                   & 100           & 0.0125    & 0     \\ \cline{2-7}
& 200                   & 4892.7                  & 4948.2                   & 200           & 0.0250    & 0      \\ \Xhline{3\arrayrulewidth}
\end{tabular}}
\label{tab:data}
\end{table}

By replacing the actual stake with percentage stake, a \cc can mitigate centralization of miners. Also, by driving baseline stakes and number of blocks from the mempool, the security and efficiency of the system can be significantly improved. However, as with any new scheme, there are application-oriented challenges address before \ep adoption by a \cc such as Bitcoin.

Bitcoin network consists of nodes running a software client (Bitcoin Core) that implements Bitcoin protocols. To incorporate \ep, the Bitcoin community would have to release a new update that implements \ep atop the existing rules set by the system. Below, we mention some other changes that may be required if Bitcoin switches to \ep.

\BfPara{Bootstrapping}
In \ep, each epoch has a dependency on the previous epoch, where miners are required to achieve PBFT-based consensus over smart contract input values. To bootstrap \ep, the first epoch needs to be independent, since it will not have a dependency on any prior execution fragment. To achieve that, we propose an intuitive solution, where any set of $k$ miners in a cryptocurrency can launch a hard fork on the \cc blockchain. Next, they can execute PBFT for the mempool state and launch the first epoch. Once the epoch is launched, the resulting blocks will be announced to the network and all peers can switch to the new fork. All the subsequent rounds can follow the same procedure outlined in this paper. After the hard fork and the bootstrap phase, any \cc can switch to \ep. 

\BfPara{Baseline Stake and Epoch Length} \label{sec:bsn} In Bitcoin, 12.5 new coins (coinbase rewards) are released when a new block is computed. The reward of miners include those newly created coins and the mining fee of transactions. As such, the baseline stake in \ep-based Bitcoin would require adding the coinbase reward as well, the baseline stake for each block($\mathcal{ST}_{i} = \mathcal{ST}_{i} + 12.5 BTC $). Moreover,in Bitcoin, the standard block size is 1MB, and there is no cap on mempool size. However, due to selective mining, there is often a transaction backlog that reduces the system efficiency. Moreover, varying the hash rate often increases the time between publication of two subsequent blocks, which at times exceeds 50 minutes. If Bitcoin employs \ep, all blocks will be published at the exact time and in each epoch all transactions will be processed.

\BfPara{Mining Record} \label{sec:epln}
In \ep, the smart contract maintains a mining record, $M(m_{j},h_{j})$, in which the history of each mining node is maintained for a future reference. This record is consulted if there is a conflict in the block auction, and preference is given to the miner with fewer blocks. In Bitcoin, the smart contract can achieve this by mapping the mining record to the IP addresses of nodes. Nodes in Bitcoin connect using IP addresses, and these IP addresses are also used as reputation indicators for the node~\cite{bitnodes_18}.

The mining record can also be maintained with respect to the accounts held by each user. However, we observe that in Bitcoin and Ethereum, a node with a single IP address can easily generate multiple accounts. This allows an adversary to split his balance into multiple accounts using the same node, and place multiple bids in each epoch.

In blockchain systems, creating new accounts is easier than creating a new IP address. For example, if the adversary wants to create $p$ unique IP addresses to influence the mining record, he must either launch $p$ unique nodes or shut down his existing node and restart with a new IP address. We note that both methods are costly. If the adversary launches $p$ unique blockchain nodes, he must download the entire blockchain ledger at each node and concurrently manage them for each auction. In contrast, if he restarts his existing machine to acquire a new IP address, his node can take up to several minutes to synchronize with the network~\cite{BitcoinCore_18,bitnodes_18}. If the next epoch starts while the adversary's nodes are synchronizing, then they will not be able to participate in the auction. Therefore, using IP addresses for mining record is a more useful approach than using accounts since switching  IP addresses can be more expensive than switching multiple accounts. 

It can be argued that a powerful adversary with more than 50\% coins can afford to launch $p$ machines, each with a unique IP address. In that case, if the adversary splits his balance among $p$ nodes then each node will have less balance than the aggregate balance of the adversary before splitting. As such, if $p$ is large, the balance maintained by each node in $p$ will be smaller. Therefore, in an auction, an honest miner with a higher balance can beat the dishonest miner in $p$ with lower balance. To avoid that, the adversary needs to know the actual balance of each honest party in the system and select the appropriate value of $p$. Since the balance of an honest party is always anonymous, the adversary cannot compute the precise value of $p$ to dominate the auction. In summary, using IP addresses to maintain a mining record helps in disincentivizing an adversary.

\BfPara{Limitations and  Future Directions}
The smart contract rules can be encoded in the software clients such as Bitcoin Core. Bitcoin Core is itself a smart contract with rules to generate, validate, and broadcast a transaction. Similarly, \ep rules can be encoded in Bitcoin Core to operate the blockchain application. Moreover, while \ep provides several features for a decentralized and fair blockchain system, it can be further improved to address the following limitations. 

The software clients of blockchain applications are vulnerable to attacks \cite{SaadCLTM19}. The security of software clients is highly critical for \ep, and within the scope of this work, we do not cover this aspect. Our simulations were executed on a prototype smart contract in which we did not face the threat of external attackers. However, part of our ongoing work is to securely integrate \ep in a modified version of Bitcoin Core and perform an end-to-end security analysis.

\ep is a multi-phased protocol involving block auction, miner selection, and mempool agreement. The message exchange in all these phases adds a time penalty to block mining. To analyze the time penalty, we use data from the Bitcoin network and try to map it on the \ep model.

On average, a Bitcoin transaction takes 2.6 seconds to reach 1000 nodes~\cite{bitnodes_18}. In some cases, the delay can be as high as 8 seconds~\cite{bitnodes_18}. If we assume $c$=1000 candidate miners, the primary replica will receive 1000 bids in parallel for the next epoch. The primary will then execute PBFT among $K$ miners to set the parameters for the next epoch. Since the average Bitcoin mempool size is $\approx$6--10 times the block size, therefore, we can assume $K$=10 miners. Prior work~\cite{SukhwaniMCTR17} shows that in a network size of 10 nodes, PBFT takes $\approx$17 milliseconds. Therefore, the PBFT execution will be quick and the bottleneck will be created the bid transmission by $c$ to the primary replica in $K$. Logically, if the candidate miners are more than 1000 (\ie 10K in Bitcoin~\cite{bitnodes_18}), the bid transmission can be delayed by more than 8 seconds. As a result, block mining will also be delayed. 

In some cryptocurrencies such as Bitcoin and Litecoin, this delay can be easily tolerated since their block mining time is 10 minutes and 2.5 minutes, respectively. However, in Ethereum the block mining time is $\approx$20 seconds, and delay caused by the auction in \ep may not be tolerable. We acknowledge this limitation and note that the constraint is due to the network size and block mining policy. If the Ethereum network switches to \ep, it will have to increase the block mining time or limit the network size. 

Another limitation of our work is the marginal sacrifice on the fault tolerance of the system. While the smart contract can prevent against the majority attacks, however, the key security bottleneck is PBFT's fault tolerance prior to the execution of the smart contract. PBFT requires at least 70\% miners to behave honestly in an epoch. In other words, the system can only tolerate up to 30\% malicious replicas. If an adversary controls more than 30\% of the miners in an epoch, then \ep may not halt due to diverging mempool views. Therefore, the fault tolerance in \ep is low compared to the other PoW and PoS-based cryptocurrencies, where it is 50\%. An alternative would be to eliminate PBFT from \ep. However, as a requirement, we need a consensus over the mempool size. In a synchronous environment, as originally considered in Bitcoin \cite{nakamoto2008bitcoin}, all nodes have the same mempool state at any given moment. In practice, \cc networks are asynchronous~\cite{SaadCLTM19}, and the mempool views may not be consistent at all times. To achieve consensus over the mempool, we incorporated PBFT consensus protocol. 

Finally, another limitation of \ep is the marginal sacrifice of anonymity due to historical dependency on IP addresses. In \ep, the smart contract maintains a history of each miner through their IP addresses. Since blockchain in \ep is a public ledger, anyone from outside the network can query and obtain the IP addresses that have mined most blocks. Accordingly, the balance of the IP address can also be calculated. Although the IP addresses in Bitcoin and Ethereum nodes are currently public~\cite{bitnodes_18}, it is perhaps worthwhile to increase the network anonymity by masking those addresses. Our future work involves using techniques to replace IP addresses with a new identity scheme in the mining repository. Moreover, besides \ep, if the IP usage is replaced by secure yet anonymous identity schemes, then it would benefit existing blockchain applications in general by hardening their security against routing attacks \cite{apostolaki2017hijacking}.

\vspace{-4mm}

\section{Related Work} \label{sec:rw}
\BfPara{PoW Limitations} Energy inefficiency of PoW-based blockchain applications has been significantly highlighted before \cite{BahriG18,Dwyer14,GiungatoRTT17,SYMITSI2018127}. Dwyer and Malone \cite{Dwyer14} firt observed the energy intensive computation performed by Bitcoin miners, and postulated it to become a major problem upon Bitcoin expansion. Bahri and Girdzijauskas \cite{BahriG18} analyzed the energy consumption of Bitcoin to achieve its core consensus protocol. Their estimates show that Bitcoin  consumes 39.5 TWh of electricity annually, on average. Harald Vranken \cite{VRANKEN20171} looked into the sustainability of Bitcoin and blockchains by performing a resource profiling of major PoW-based blockchain applications. His work also attributes the growing energy footprint of the Bitcoin \cc to the endogenous PoW requirements. 

\BfPara{PoS Limitations} Although PoS addresses the growing energy problems of PoW \cite{Spasovski17,BartolettiLP17,FanZ17}, it is vulnerable to a series of attacks and introduces unfairness in the system. The phenomenon of ``{\em the rich gets richer}'' in mining has been reported \cite{ZhengXDCW17,Ren14,BentovGM16}. In PoS, the baseline stakes create a partition in the network in which the number of peers above the baseline continuously benefit from the fee rewards. This keeps increasing the threshold for baseline stake and the margins of partition between the wealthy nodes and the rest of the network. Zheng \etal noticed this limitation of PoS, and highlighted the need for a new PoS algorithm. Towards that, Kiayias \etal~\cite{KiayiasRDO17} performed a formal analysis of PoS-based blockchains using the PoW-based theoretical model proposed by Garay \etal~\cite{GarayKL15}. They formally specified the desirable security properties for a PoS blockchain system, and using those properties they present a model called ``Ouroboros.'' Ouroboros uses a unique reward mechanism to incentivize PoS-based systems and neutralize the selfish mining attacks. In Ouroboros, the authors use a coin-flipping technique to introduce randomness in miner selection. However, as we have shown in \autoref{tab:data}, the random selection, despite its benefits, may sacrifice fairness. To ensure fairness, it is important that the balance of each stakeholder must be above the baseline stake. In random selection (see \autoref{fig:rs}), the balance of a miner was often below the baseline stake, which can compromise the security. Therefore, specific to the \ep design, random selection of miners can be counterproductive since it sidesteps the security requirements of the baseline stake.  

In a similar context, Diant \etal~\cite{DaianPS19} presented \textsc{Snow White}, a variant of PoS-based consensus protocol that realizes a secure application of PoS in {\em permissionless} blockchain systems. Similar to the work of \cite{KiayiasRDO17}, \textsc{Snow White} also utilizes random selection of miners to achieve decentralization. However, instead of the coin-flipping method in~\cite{DaianPS19}, \textsc{Snow White} is an extension of the ``Sleepy Consensus''~\cite{PassS17} and introduces a dependency among the block headers to achieve the notion of randomness. Finally, another notable work is by Chen \etal~\cite{ChenM19}, called Algorand, which uses message-passing Byzantine Agreement to achieve consensus in large-scale distributed systems. Algorand requires significantly less energy to operate and it is currently being deployed in a few blockchain systems.  

Finally, two other notable PoS-based protocols are Delegated Proof-of-Stake (DPoS) and Supernode Proof-of-Stake (SPoS). In (DPoS), the network participants vote to elect a team of witnesses that mines block~\cite{WangLL20}. SPoS is an extension of DPoS in which supernodes are elected to mine blocks~\cite{blockchain_community_20}. Compared to DPoS, SPoS guarantees a constant inter-arrival block time and optimized data storage. In both protocols, miners commit their stakes before mining a block. If miners misbehave, their stakes are confiscated. The key problem with this approach is that if the reward of misbehavior is greater than the stake, a miner can easily cheat, violating the fairness property. No mechanism binds stakes with fee rewards to prevent malicious behavior.  

In \ep, we overcome this problem by using the concept of baseline stake driven from the mempool. The mempool allows us to precisely calculate the baseline stakes for each block in an epoch and set the lower bound commitment. As a result, the miner's stake is always greater than or equal to the baseline stake for each block. If a miner cheats, all victims can be compensated. This feature enables fairness in \ep and distinguishes it from both SPoS and DPoS. 

Another limitation of DPoS and SPoS is that they cannot work in an asynchronous network. For instance, in SPoS, and all witnesses are expected to have a consistent network view while mining blocks. However, as we extensively discuss in section 3.5, when the blockchain network size increases, the network can exhibit asynchrony which eventually leads to inconsistency among miners. Both DPoS and SPoS ignore the effect network asynchrony. In \ep, we address asynchrony by using the PBFT consensus protocol that synchronizes miners during epochs.

\BfPara{Hybrid Designs} To address the limitations these consensus schemes, hybrid approaches have been adopted to converge their benefits and reduce the attack surface. Duong \etal \cite{DuongCFZ18} proposed {\em TwinsCoin}; a secure and scalable hybrid blockchain protocol. Bentov \etal \cite{BentovLMR14} proposed a hybrid blockchain protocol called  Proof-of-Activity (PoA) which upgrades the defense measures of blockchains with low penalty on network communications and storage space. Some notable attempts have been made to secure blockchains against centralization and majority attacks \cite{DuongCZ18,LiABK17,SpasovskiE17,rocket2018snowflake}. Despite these commendable efforts, there is still need for a practical solution that can be applied to existing cryptocurrencies to enable them to easily transition to PoS. Our proposed model addresses these limitations, and presents a solution that can be used in the existing blockchain systems with various desirable properties.  
\vspace{-4mm}
\section{Conclusion} \label{sec:con}
Due to the massive energy footprint and great computation requirements, PoW-based blockchain applications are becoming infeasible. In contrast, the popular alternative known as the Proof-of-Stake (PoS) suffers from network centralization and unfairness.  In this paper, we have introduced an extended form of PoS, termed as \ep, which  introduces fairness in the blockchain network and resists centralization. We introduce a smart contract that runs atop the blockchain and carries out a blind block auction. The smart contract applies policies to extend mining opportunities to a wider set of network peers, and ensures fair reward distribution. The execution of the smart contract is facilitated by miners that run PBFT-based consensus over the mempool state. We show, by simulation, that \ep meets its objectives and can be applied to existing blockchain systems, such as Bitcoin.   

\vspace{-3mm}
 \balance
\bibliographystyle{IEEEtran}
\bibliography{Ref/ref.bib,Ref/conf.bib}

\end{document}